\documentclass[aps,amsfonts,prb,twocolumn,showpacs,eqsecnum]{revtex4-1}
\usepackage{epsfig,amsmath,amssymb,bm,epsf,graphics,psfrag,verbatim,subfigure}
\usepackage{slashed}
\def\confenergy{U} 
\def\enddist{P} 
\def\plength{L} 
\def\pwidth{w} 
\def\qpos{q} 
\def\qtime{t}
\def\qTime{{{\cal T}}} 
\def\qham{H} 
\def\pdensity{\rho} 
\def\newpdensity{Q} 
\def\newx{s}
\def\mdensity{\bar{\rho}} 
\def\onebody{\Phi}  
\def\dummyxprime{\bar{x}} 
\def\newdummyxprime{\bar{s}} 
\def\sgn{\mathop{\rm sgn}}
\def\gap{\gamma} 
\def\disp{\delta}

\begin{document}

\title{Directed-polymer systems explored via their quantum analogs: \\ topological constraints and their consequences}

\author{D. Zeb Rocklin$^{1}$, Shina Tan$^{2}$ and Paul M. Goldbart$^{2}$}

\affiliation{
$^1$Department of Physics, University of Illinois at Urbana-Champaign,
1110 West Green Street, Urbana, Illinois 61801, USA}

\affiliation{
$^2$School of Physics, Georgia Institute of Technology,
Atlanta, Georgia 30332, USA}

\date{\today}

\begin{abstract}
The equilibrium statistical mechanics of classical directed polymers in 2~dimensions is well known to be equivalent to the imaginary-time quantum dynamics of a 1+1-dimensional many-particle system, with polymer configurations corresponding to particle world-lines. This equivalence motivates the application of techniques originally designed for one-dimensional many-particle quantum systems to the exploration of many-polymer systems, as first recognized and exploited by P.-G.~de~Gennes [J.\ Chem.\ Phys.\ {\bf 48\/}, 2257 (1968)]. In this low-dimensional setting interactions give rise to an emergent polymer fluid, and we examine how topological constraints on this polymer fluid (e.g., due to uncrossable pins or barriers) and their geometry give rise to strong, entropy-driven forces. 
In the limit of large polymer densities, in which a type of mean-field theory is accurate, we find that a point-like pin causes a divergent pile-up of polymer density on the high-density side of the pin and a zero-density region (or gap) of finite area on the low-density. In addition, we find that the force acting on a pin that is only mildly displaced from its equilibrium position is sub-Hookean, growing less than linearly with the displacement, and that the gap created by the pin also grows sublinearly with the displacement. By contrast, the forces acting between multiple pins separated along the direction preferred by the polymers are super-Hookean. These nonlinear responses result from effective long-ranged interactions between polymer segments, which emerge via short-ranged interactions between distant segments of long polymer strands.
In the present paper, we focus on the case of an infinitely strong, repulsive contact interaction, which ensures that the polymers completely avoid one another. In a companion paper, we consider the effects of a wider set of inter-polymer interactions.
\end{abstract}

\maketitle

\section{Introduction}

The ensemble of configurations of a set of directed, one-dimensional objects embedded in $(D+1)$-dimensional space can be envisioned as the ensemble of worldlines of a corresponding set of quantum-mechanical point particles evolving in time in $D$ spatial dimensions.  In particular, a standard mapping relates the classical canonical-ensemble equilibrium statistical mechanics of the set of directed one-dimensional objects to the imaginary-time evolution of the state of the corresponding set of point particles.  This mapping was introduced and exploited by de~Gennes~\cite{pdg} in order to shed light on the equilibrium structure of directed fibrous polymers that are confined to two dimensions, thus providing a scheme for accounting, nonperturbatively, for strong local polymer-polymer interactions that serve to prohibit configurations in which polymers cross.  For de~Gennes, this prohibition is accomplished by asserting that the quantum particles are identical fermions (in his case free, noninteracting, and subject to periodic spatial boundary conditions), and are therefore subject to the Pauli exclusion principle.

In a parallel development, a suite of powerful techniques---specifically, Bethe's Ansatz, bosonization, and quantum hydrodynamics---have been developed to address the quantum mechanics of one-dimensional systems of many interacting particles or spins.  The aim of the present paper and a companion~\cite{ref:DZRcompanion} one is to employ these advances in quantum many-body (QMB) physics, together with the de~Gennes analogy between the quantum-fluctuating, many-particle system and the classical, thermally fluctuating directed-polymer system, to uncover new information about the equilibrium structure of systems comprising polymers that are either rigorously prohibited from passing through one another (i.e., noncrossing) or subject to other interactions, such as energetic penalizations of crossings, or systems that allow for the presence of distinct species of polymers.  In addition, we apply these advances in technique to determine the equilibrium forces acting on particles included in interacting polymer systems that serve to exclude the polymers from certain spatial regions, as well as the effective forces that act between such particles as a result of their exclusion of polymers.  A global theme of the present work is that, due to the reduced dimensionality of the polymer system, interactions dramatically influence the structure and correlations that characterize the polymer system, and do so both, as we shall see, in topologically---and also geometrically---rich settings.  This is a lesson already well known in the quantum-particle domain.  Although we focus in this paper on polymer systems, our treatment applies to other two-dimensional statistical systems involving linelike degrees of freedom, such as wandering steps edges on crystal surfaces~\cite{twd}, dynamically growing interfaces in the Kardar-Parisi-Zhang universality class~\cite{kulkarni}, and vortex lines in planar type-II superconductors~\cite{kafri}.

The present paper is organized as follows. In Sec.~\ref{sec:partitionfcn} we introduce the two-dimensional interacting directed polymer system. We also describe its mapping to a one-dimensional quantum analog, and address the statistical observables that may be derived via this mapping. In Sec.~\ref{sec:toposection} we describe how to impose and analyze topological constraints on the polymer system, and we discuss in detail the resulting effects on the free energy and structure of the polymer system. In Sec.~\ref{sec:conclusion} we summarize our results and provide some conclusions. In a companion paper~\cite{ref:DZRcompanion}, we consider more general classes of polymer interactions. For such systems, we analyze the resulting interpolymer correlations as well as the response of the polymer system to impurities such as free or fixed lines or particles, which are not topological in character. In addition, we discuss the application of the technique of bosonization, familiar from quantum many-body physics, as a tool for characterizing the universal behavior of directed polymer systems.

\section{Directed 2D classical equilibrium polymers and evolving 1D quantum particles}
\label{sec:partitionfcn}

\subsection{Directed 2D classical polymers in thermal equilibrium}

\begin{figure}[hh]
\centerline{
\includegraphics[width=.48\textwidth]{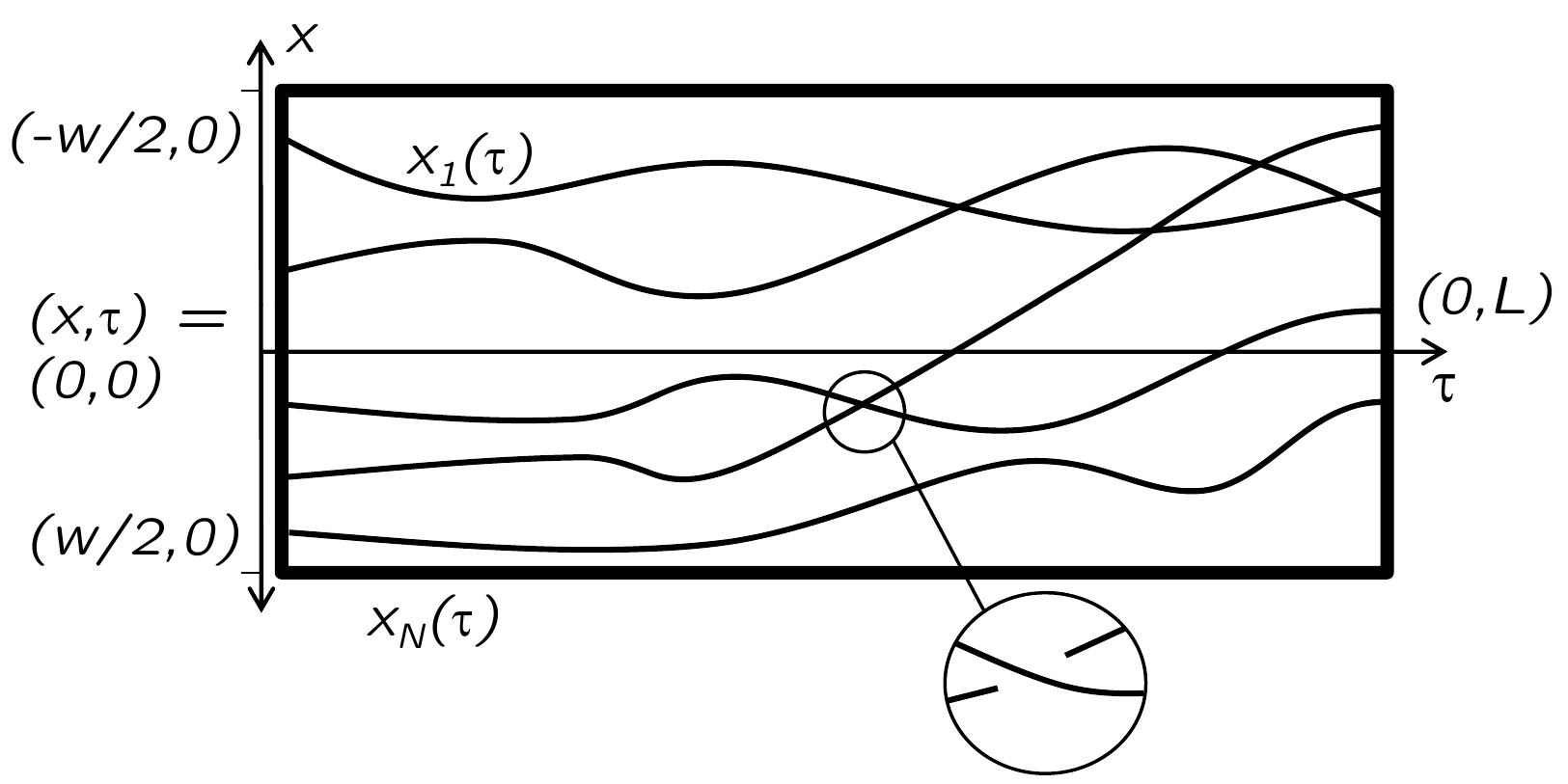}}
\caption{
The paths $\{x_n(\tau)\}$ describe a possible configuration of the directed polymer system. Thermal fluctuations permit the system to adopt energetically disfavored configurations. When polymers appear to intersect in the $(x,\tau)$ plane, in reality one crosses over the other by exploiting the presence of a third dimension.
}
\label{fig:crossingdiagram}
\end{figure}

We consider a system of $N$ two-dimensional (2D) directed polymers, indexed by $n=1,\ldots,N$, that are noncrossing~\cite{knotting}.  The configuration of the $n^{\rm th}$ polymer is described by $x_n(\tau)$, where $\tau$ is the coordinate along the directed axis of the system (which we call the longitudinal direction), and $x_n$ gives the location of the $n^{\rm th}$ polymer in the perpendicular direction (which we call the lateral direction); see Fig.~\ref{fig:crossingdiagram}. 
We take the energy cost of the deflections of the polymers from the longitudinal direction to be
\begin{equation}
\label{eq:bendingeq}
\frac{A}{2}\sum_{n=1}^{N}
\int_0^{\plength}d\tau\,(\partial_\tau x_{n})^2,
\end{equation}
where $L$ is the extent of system in the longitudinal direction, and $A$ is the bending energy per unit length, which penalizes configurations for straying from the preferred (i.e., longitudinal) direction. The polymers must be stiff in a way which we will define later to permit us to neglect higher-order terms in this expression.
In addition to the bending energy, we include an interaction $V$ between the polymers, which we take to be translationally and parity invariant, and sufficiently short-ranged that it may be taken to operate only between monomers (i.e., polymer segments) having common $\tau$ coordinates.  In fact, we shall often take the interaction to be purely local, in which case it would take the form
$V\big(x_{n}(\tau)-x_{n^{\prime}}(\tau)\big)
=c\,\delta\big(x_{n}(\tau)-x_{n^{\prime}}(\tau)\big)$,
where $\delta(x)$ is the one-dimensional Dirac delta function. Thus, we arrive at the following energy functional $\confenergy$ of a configuration $\{x_{n}(\cdot)\}_{n=1}^{N}$ of the polymer system:
\begin{eqnarray}
&&\confenergy\left[\{x_n(\cdot)\}\right]=
\frac{A}{2}\sum_{n=1}^{N}
\int_0^{\plength} d\tau\,
\big(
\partial_\tau x_{n}(\tau)
\big)^2
\nonumber
\\
&&\qquad
+\frac{1}{\plength}\sum_{n=1}^{N}
\int_0^{\plength} d\tau\,
\onebody\big(x_{n}(\tau)\big)
\nonumber
\\
&&\qquad
+\frac{1}{\plength}
\sum_{1\le n<n^{\prime}\le N}
\int_0^{\plength} d\tau\,
V\big(x_{n}(\tau)-x_{n^{\prime}}(\tau)\big),
\end{eqnarray}
in which we have also included an external (or one-body) potential term $\onebody$.
According to this model, in the absence of polymer-polymer interactions or external potentials the polymer configurations have a thermal distribution that is Gaussian, in the sense that increments in their deflections
[i.e., $x_{n}(\tau+\delta\tau)-x_{n}(\tau)$]
are independent Gaussian random variables having mean zero and variance $\delta \tau k_B T/A$, where $T$ is the system temperature and $k_B$ is Boltzmann's constant, which we generally set to unity via a suitable choice of units.  We note that we shall not be considering the {\it dynamics\/} of the polymer system, so we do not need to take note of the kinetic energy of the polymer system.  Additionally, we characterize the system via conditions on the configurations of the polymers at their ends, via the distributions
$\enddist^{i}(\{x_{n}\})$ and
$\enddist^{f}(\{x_{n}\})$,
which respectively give the probability densities for the configurations $\{x_{n}\}$
of the polymer ends at $\tau=0$ and $\tau=\plength$.

We take the polymer system to be at thermal equilibrium at inverse temperature $\beta$.  Thus, we have for the canonical ensemble partition function~\cite{gibbsdenominator}
\begin{eqnarray}
&&Z[\enddist^{f},\enddist^{i}]=
\int
d\{X_{n}^{f}\}\,\enddist^{f}(\{X_{n}^f\})\,
d\{X_{n}^{i}\}\,\enddist^{i}(\{X_{n}^i\})\,
\nonumber\\
&&\qquad\qquad\quad\times
\int\limits_{\{x_{n}(0)=X_{n}^{i}\}}^{\{x_{n}(L)=X_{n}^{f}\}}
\!\!\!\!\!\!\!\!\!\!\!
\mathcal{D}\left[\{x_{n}(\cdot)\}\right]\,
{\rm e}^{-\beta\confenergy\left[\{ x_n(\cdot) \} \right]},
\end{eqnarray}
which depends functionally on $\enddist^{i}$ and $\enddist^{f}$.
Here, the measures are defined via
$d\{X_{n}^{f}\}\,d\{X_{n}^{i}\}\equiv
\prod_{n=1}^{N}dX_{n}^{f}\,dX_{n}^{i}$
and
$\mathcal{D}\left[\{x_{n}(\cdot)\}\right]\equiv
\prod_{n=1}^{N}
\mathcal{D}\left[x_{n}(\cdot)\right]$.
To complete the definition of this multiple path integral we also need to impose some form of lateral boundary conditions on the polymer configurations.  We return to this point in Sec.~\ref{sec:noncrossing}.

As for the thermal expectation value $\langle O\left[\{x_{n}(\cdot)\}\right]\rangle$ of a generic observable (i.e., a functional of the polymer configuration) $O\left[\{x_{n}(\cdot)\}\right]$, this is given by
\begin{eqnarray} \label{eq:observable}
&&\langle O \rangle=
Z[\enddist^{f},\enddist^{i}]^{-1}
\!\!\int
d\{X_{n}^{f}\}\,\enddist^{f}(\{X_{n}^f\})\,
d\{X_{n}^{i}\}\,\enddist^{i}(\{X_{n}^i\})\,
\nonumber\\
&&\qquad\times
\int\limits_{\{x_{n}(0)=X_{n}^{i}\}}^{\{x_{n}(L)=X_{n}^{f}\}}
\!\!\!\!\!\!\!\!!\!\!
\mathcal{D}\left[\{x_{n}(\cdot)\}\right]\,
{\rm e}^{-\beta\confenergy\left[\{ x_i(\cdot) \} \right]}\,
O\left[\{x_{n}(\tau)\}\right].
\end{eqnarray}
As we have noted above and shall see below, owing to the low dimensionality of this thermally fluctuating polymer system, even short-ranged interactions produce qualitative alterations of the structure and correlations that it exhibits, relative to those exhibited by its noninteracting counterpart.  Moreover, even interactions that are weak, microscopically, are fundamentally nonperturbative, in that they induce correlations that are long-ranged.

\subsection{Mapping to 1D quantum particles}

Let us turn now to the consideration of a one-dimensional quantum system of $N$ nonrelativistic particles each of mass $m$ and having coordinates
$\{\qpos_{n}\}_{n=1}^{N}$,
subject to a one-body interaction $\Phi(q_n(t))$ and to a (translationally and parity invariant) two-body interaction
$V(\qpos_{n}-\qpos_{n^{\prime}})$.
For this system, and introducing the (unsymmetrized) simultaneous particle-position eigenkets
$\vert\{\qpos_{n}^{i}\})$, a matrix element of the imaginary-time propagator
$(\{X_{n}^{f}\}\vert{{\rm e}^{-\qham\qTime/\hbar}}\vert\{X_{n}^{i}\})$
can be expressed as the following Feynman integral over paths $\{\qpos_{n}(t)\}$;
(see, e.g., Ref.~\cite{feynman}):
\begin{eqnarray}
&&(\{X_{n}^{f}\}\vert{{\rm e}^{-\qham\qTime/\hbar}}\vert\{X_{n}^{i}\})
\nonumber\\
&&\qquad\qquad\qquad=
\!\!\!\!\!\!\!\!\!\!\!
\int\limits_{\{\qpos_{n}(0)=X_{n}^{i}\}}^{\{\qpos_{n}(\qTime)=X_{n}^{f}\}}
\!\!\!\!\!\!\!\!!\!\!
\mathcal{D}\left[\{\qpos_{n}(\cdot)\}\right]\,
{\rm e}^{-S_{E}\left[\{\qpos_n(\cdot)\}\right]/\hbar},
\end{eqnarray}
where the Euclidean action $S_{E}$ is given by
\begin{eqnarray}
\label{eq:quantumaction}
&&S_{E}=
\int_0^{\qTime}
d\qtime\,\Big\{
\sum_{n=1}^{N}
\frac{m}{2}
\left(\partial_\qtime\qpos_{n}\right)^2+\sum_{n=1}^N \Phi(q_n)
\nonumber\\
&&\qquad\qquad\qquad
+
\sum_{1\le n<n^{\prime}\le N}
\!\!\!\!\!\!\!
V\big(\qpos_{n}(\qtime)-\qpos_{n^{\prime}}(\qtime)\big)
\Big\}.
\end{eqnarray}
The \lq\lq non-Lagrangian\rq\rq\ sign of the interaction term in Eq.~(\ref{eq:quantumaction}) and the terminology of the Euclidean action reflect the fact that we are considering imaginary-time propagation, in which the paths of the Feynman integral are  referred to as imaginary-time world-lines.  As usual, the propagator can be used to construct the transition amplitude between generic initial and final quantum states $\vert{\Psi^{i}}\rangle$ and $\vert{\Psi^{f}}\rangle$:
\begin{eqnarray}
&&\langle{\Psi^{f}}\vert
{{\rm e}^{-\qham\qTime/\hbar}}
\vert{\Psi^{i}}\rangle=
\int
d\{X_{n}^{f}\}\,
d\{X_{n}^{i}\}\,
\langle{\Psi^{f}}\vert\{X_{n}^{f}\})
\nonumber\\
&&\qquad\qquad
\times
(\{X_{n}^{f}\}\vert{{\rm e}^{-\qham\qTime/\hbar}}\vert\{X_{n}^{i}\})
(\{X_{n}^{i}\}\vert{\Psi^{i}}\rangle.
\end{eqnarray}

In order to relate this quantum-mechanical system to the classical polymer systems, we make the following identifications.  We match the wave functions with the probability distributions, i.e., we choose
\begin{subequations}
\label{eqs:matching}
\begin{eqnarray}
(\{X_{n}\}\vert{\Psi^{i}}\rangle^{\phantom{\ast}}&=&P^{i}(\{X_{n}\}),
\\
(\{X_{n}\}\vert{\Psi^{f}}\rangle^{\ast}&=&P^{f}(\{X_{n}\}).
\end{eqnarray}
\end{subequations}
Note that the identification is not between classical and quantal probability distributions but, rather, between classical probability distributions and quantal wave functions.  Thus, to be appropriate, the wave functions should be restricted to being real, non-negative, and integrating to unity.  The equivalence between the quantal and classical problems is completed by adding the following identifications:
\begin{subequations} \label{eq:mappings}
\begin{eqnarray}
\qTime&=&\hbar\beta,\\
\qtime&=&\qTime\tau/\plength,\\
\qpos_{n}(\qtime)&=&x_{n}(\tau),\\
m&=&\hbar^{2}\beta^{2}A/\plength.
\end{eqnarray}
\end{subequations}
In order to maintain the analogy to the polymer system, we refer to the 1D quantum system as having a \emph{width} rather than a length, and denote the widths of both systems by $\pwidth$.  Then we have the result that is central to this paper, viz., that the quantal matrix element of the imaginary-time evolution operator is equal to the classical partition function of the polymer system:
\begin{eqnarray}
\langle{\Psi^{f}}\vert{{\rm e}^{-\qham\qTime/\hbar}}\vert{\Psi^{i}}\rangle
=
Z[P^{f},P^{i}].
\end{eqnarray}
In particular, the imaginary-time worldlines of the quantum particles correspond to the configurations of the directed polymers, and the quantum fluctuations of the particle system (the strength of which is governed by $\hbar$) correspond to the thermal fluctuations of the polymer system (the strength of which is governed by $1/\beta$).

In order to apply many of the techniques of QMB physics it is useful to impose a choice of quantum statistics upon the initial and final quantum states $\vert{\Psi^{i}}\rangle$ and $\vert{\Psi^{f}}\rangle$.  The condition that the initial and final wave functions
$(\{X_{n}\}\vert{\Psi^{i}}\rangle$ and $(\{X_{n}\}\vert{\Psi^{f}}\rangle$
be non-negative, so that they may match the polymer end distributions, Eqs.~(\ref{eqs:matching}), precludes the direct choice of fermionic statistics (although, as we shall see, such statistics can be applied indirectly, following a Jordan-Wigner--type transformation).
Thus, we choose to consider situations in which the polymer endpoint distributions $P^{i}$ and $P^{f}$ are each symmetric functions---respectively under the exchange of
initial endpoints amongst themselves and
final   endpoints amongst themselves---from
which it follows that the initial and final quantum end states be symmetric under particle exchange, and hence that they describe identical \emph{bosons}~\cite{fermistates}.  The quantum Hamiltonian that corresponds to the Euclidean action~(\ref{eq:quantumaction}) is invariant under particle exchange (i.e., $P H P^\dagger = H$ for all pairwise particle exchange operators $P$), and therefore under the imaginary-time evolution described by ${\rm e}^{-H \tau/\hbar}$ bosonic many-body states remain bosonic.

One valuable notion made accessible and sharpened via the tools of one-dimensional quantum many-body systems (such as bosonization, quantum hydrodynamics, and Bethe-Ansatz-rooted methods) is that of the {\it emergent directed-polymer liquid\/} (cf.~Kafri \textit{et al}.~\cite{kafri}).  This state is a classical analog of the Luttinger-Tomonaga liquid, which can be exhibited by interacting QMB systems in one dimension, and is qualitatively distinct from, e.g., the Landau liquid state of many-fermion systems, i.e., a state that can be exhibited by such systems in higher dimensions.  Thus, we shall see, e.g., that the manner in which density correlations decay spatially in the emergent directed-polymer liquid resembles the space-time decay of superfluid fluctuation correlations in one-dimensional interacting QMB systems.

\subsection{Eigenfunction expansion of the imaginary-time propagator}
\label{sec:eigfcn}

Due to the association of polymer configurations with particle paths, discussed in the previous subsection, it is the path-integral (i.e., covariant) formulation of the quantization of the system that is the one most clearly associated with the physical degrees of freedom of the fluctuating polymer system.  However, the quantum mechanics of the same particle system can also be formulated in terms of the time-dependent Schr{\"o}dinger equation (i.e., via canonical quantization).  Thus, it is straightforward to ascertain that the propagator can be expressed in terms of the expansion
\begin{eqnarray}
&&(\{x_{n}^{f}\}\vert{{\rm e}^{-\qham\qtime/\hbar}}\vert\{x_{n}^{i}\})
\nonumber\\
&&\qquad\quad=
\sum\nolimits_k e^{-E_k\qtime/\hbar}\,
\psi_k(\{x_{n}^{f}\})\,
\psi_k^\ast (\{x_n^{i}\})
\end{eqnarray}
over the exact normalized eigenfunctions $\{\psi_k\}$
and corresponding energy eigenvalues $\{E_k\}$, with many-body quantum numbers $k$,
of the associated many-body quantum Hamiltonian
\begin{eqnarray}
{H} = \sum_{n=1}^{N}
\frac{{p}_n^2}{2 m}\quad+
\sum_{n=1}^{N}
\onebody\big(x_{n}(\tau)\big)\\+
\sum_{1\le n<n^{\prime}\le N}V({x}_n-{x}_{n^{\prime}}).
\end{eqnarray}

An important special case is made evident via this expansion.  Suppose one is concerned with a statistical-mechanical expectation value involving polymer observables all taken at a single value of the longitudinal coordinate $\tau$ and, moreover, obeying
$(\qtime,\qTime-\qtime)\gg\hbar/\Delta E$,
where $\Delta E$ is the spacing between the ground many-body state $\vert \psi_{gs}\rangle$ of the QMB system and its  first excited state.  In this case, it can be adequate to retain only the ground state in the eigenfunction expansion, i.e., to take
\begin{eqnarray}
&&{\rm e}^{-\qham\qtime/\hbar}=
\sum\nolimits_{k}
\vert\psi_{k}\rangle\,
{{\rm e}^{-E_{k}\qtime/\hbar}}\,
\langle\psi_{k}\vert
\nonumber\\
&&\phantom{{\rm e}^{-\qham\qtime/\hbar}}\approx
\vert\psi_{gs}\rangle\,
{{\rm e}^{-E_{gs}\qtime/\hbar}}\,
\langle \psi_{gs}\vert,
\end{eqnarray}
a situation referred to as ground-state dominance.  In particular, within the ground-state dominance approximation and far from the system ends, the equilibrium expectation value of the polymer density
\begin{eqnarray}
\big\langle\sum_{n}\delta\big(x-{x}_{n}(\tau)\big)\big\rangle
\end{eqnarray}
is given, as a function of lateral
position $x$, by the quantum-mechanical ground-state expectation value of the density operator $\sum_{n}\delta(x-\hat{q}_{n})$,
a result that holds regardless of the longitudinal boundary conditions on the polymer configurations.  Thus, properties deep in the longitudinal interior of a long system (i.e., one for which $\qTime\gg \hbar/\Delta E$) are associated with the ground-state properties of the quantum system.

\subsection{Noncrossing polymers}
\label{sec:noncrossing}

\begin{figure}[hh]
\centerline{
\includegraphics[width=.48\textwidth]{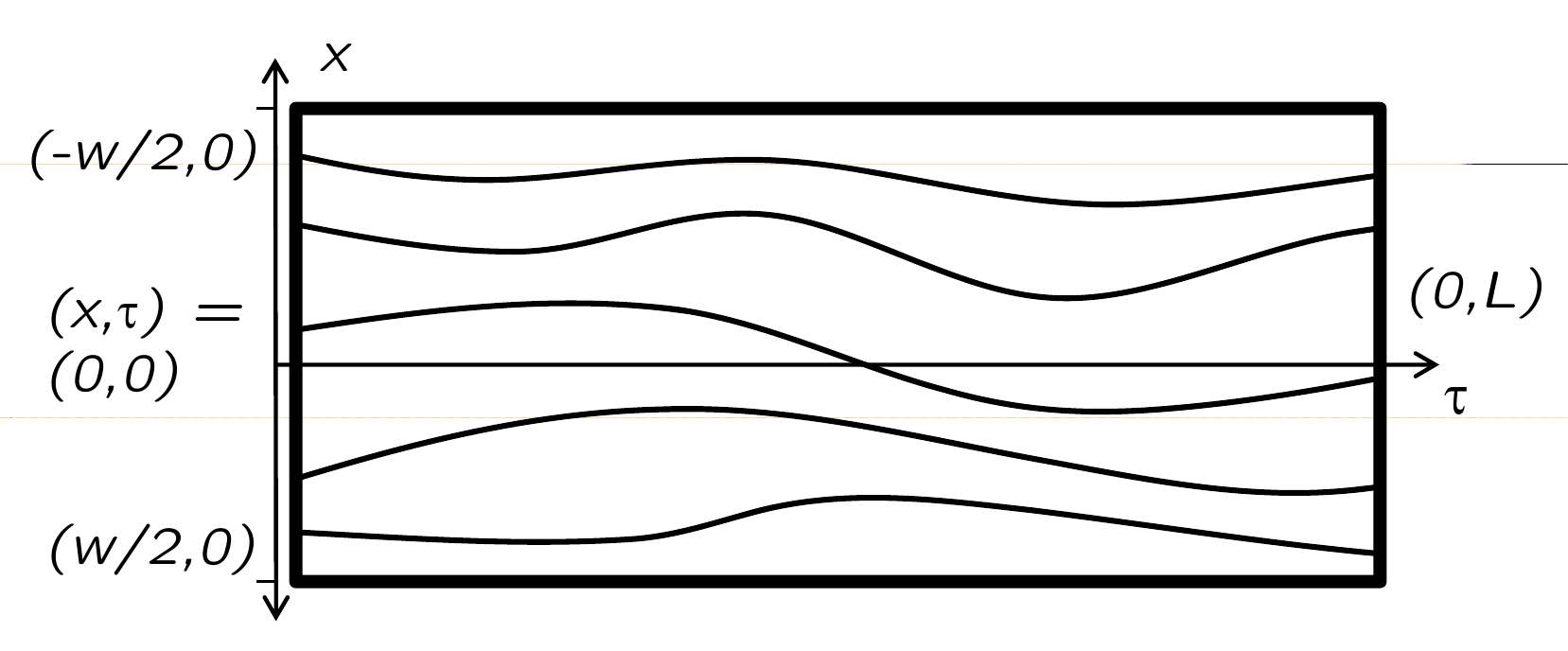}}
\caption{\label{phase}
Snapshot of a configuration for the case of noncrossing polymers.
Such polymers are not permitted to cross one another, but are otherwise noninteracting.
}
\label{fig:noncrossingdiagram}
\end{figure}

Our focus is on systems comprising strictly \emph{noncrossing} polymers.  Such systems do not adopt configurations in which any polymer crosses any other but, beyond this important element, they are not subject to any polymer-polymer interactions; see Fig.~\ref{fig:noncrossingdiagram}.
As we shall see in a companion paper~\cite{ref:DZRcompanion}, QMB physics techniques enable the study of polymer systems having a wide range of interactions; we shall show there that in the presence of such interactions many of the results obtained in the present paper will continue to hold, at least qualitatively.
To enforce noncrossing, the polymers are taken to feature an infinitely strong excluded-volume effect, so that the partition function contains only those paths for which
$x_{n}(\tau)\neq x_{n'}(\tau) $ for all $ n\neq n'$.
This restriction can be enforced for the polymer system via the inclusion of the interaction term
\begin{eqnarray}
\label{eq:hcBoson}
V\big(x_{n}(\tau)-x_{n'}(\tau)\big)=c\,\delta\big(x_{n}(\tau)-x_{n'}(\tau)\big),
\end{eqnarray}
with $c/LA\rightarrow\infty$.  The corresponding quantum system can therefore be taken to comprise many identical particles, bosonic in their quantum statistics and subject to an interparticle interaction having the same form, Eq.~(\ref{eq:hcBoson}).  This system is known as the hard-core (or impenetrable) point-like boson model.  (A finitely strong  interpolymer repulsion corresponds to a quantum Lieb-Liniger system~\cite{ll}.)\thinspace\ We mention that, despite its short range, this interaction is strong, and we may therefore expect the qualitative behavior of noncrossing polymer systems to be replicated in systems having more general interactions.

\subsubsection{From bosons to fermions}

It was demonstrated by Girardeau~\cite{girardeau} that any system of interacting, one-dimensional, bosonic particles for which the wave functions vanish when pairs of particles coincide in space can be mapped to an equivalent system of fermionic particles, subject to the same interaction $V(x_i-x_j)$ whenever $x_i \neq x_j$.  Girardeau's mapping is particularly useful in the case of hard-core boson systems, as these can be mapped to systems of \emph{free} fermions.  Thus, as first noted and exploited by de Gennes~\cite{pdg}, the noncrossing condition on polymers can be accounted for entirely by the quantum statistics of fermions, without the need to include any interaction term. 
We mention that Girardeau's mapping preserves the modulus of the quantum wave function in the \emph{position} basis; however, the forms of the bosonic and fermionic momentum-space wave functions are not preserved.  
This means that the local density of polymers is correctly described under the mapping but, e.g., 
quantities involving the slopes of polymer configurations---the analogs of the momenta of the quantum particles---are not.
De~Gennes applied this free fermion picture of two-dimensional noncrossing fibrous polymers to describe the structure of such systems.  In particular, he showed that there is a logarithmic divergence in the \lq\lq x-ray form factor\rq\rq\ (i.e., the longitudinally-averaged correlator between lateral
Fourier components of the density fluctuations of the polymer system), at a lengthscale associated with the Fermi momentum of the quantum system (or, equivalently, the mean lateral inter-polymer spacing).

\subsubsection{Ground state and ground-state dominance}

As noted in Sec.~\ref{sec:eigfcn}, the ground state of the quantum Hamiltonian plays a key role in the behavior of the polymer system over long distances.  In order to make use of this idea, we now obtain the ground-state wave function for the many-hard-core boson system, subject to hard-wall boundary conditions, in a form that is convenient for the subsequent analysis.  To do this, we begin with the ground-state wave function $\psi_{gs}^{p}(\{x_n\})$ of a system of $N$ hard-core bosons on a ring of circumference $\pwidth$, subject to {\it periodic\/} (rather than hard-wall) boundary conditions (see e.g. \cite{girardeau}); this is given by
\begin{eqnarray}
&&
\psi_{gs}^{p}(\{x_n\})=
\frac{2^{N(N-1)/2}}{\pwidth^{N/2}\sqrt{N!}}\,
\nonumber\\
&&\qquad\qquad\times
\prod_{1\le n<n'\le N}
\big\vert\sin
\frac{\pi}{\pwidth}
\big({x_n-x_{n'}}\big)\big\vert.
\end{eqnarray}
Such boundary conditions are appropriate for a system of polymers that lie on a cylindrical surface and are directed along the cylinder axis.  Our aim, however, is to consider a system of polymers that are confined to a strip with hard-wall boundary conditions, and this system corresponds to a quantum system also subject to hard-wall boundary conditions.  
To that end, we give the ground state wave function of a system of $N$ bosons subject to vanishing boundary condtions at $x_n = \pm \pwidth/2$ (see Appendix~\ref{sec:ellipticals}):

\begin{eqnarray}
\label{eq:gswavefcn}
\psi_{gs}(\{x_n\})
=
\frac{2^{N^{2}/2}}{\pwidth^{N/2} \sqrt{N!}}
\bigg(
\prod_{n=1}^{N}
\cos \frac{ \pi x_{n}}{\pwidth}
\bigg) \nonumber \\
\times \bigg(
\prod_{1\le n<n^{\prime} \le N}
\!\!\!\!\!\!
\big\vert\sin \frac{\pi x_{n}}{\pwidth}-
\sin \frac{\pi x_{n^\prime}}{\pwidth}
\big\vert
\bigg).
\end{eqnarray}

\noindent The wave function $\psi_{gs}$ reflects the inter-polymer repulsion.  Although the corresponding polymers are forbidden energetically only from actually intersecting one another, continuity and thermal fluctuations have the combined effect of \lq\lq carving out\rq\rq\ a spatial region around the polymers so that the probability of finding one polymer very near another (compared with the mean inter-polymer spacing $\pwidth/N$) vanishes as the square of the separation.  A similar effect occurs near the hard boundaries, i.e., at $x=\pm\pwidth/2$.
The preceding results pertain to infinitely strong contact interactions. However, as is known from the work of Lieb and Liniger~\cite{ll}, the physical properties of a system of bosons subject even to \emph{weak} contact interactions differ nonperturbatively from those of a system of free bosons.

Whereas the lateral {\it correlations\/} amongst the polymer segments depend additionally on the quantum-mechanical energy eigenfunctions, the thermodynamic properties of the polymer system are determined solely by the spectrum of energy eigenvalues.  The ground-state energy and energy spacing to the first excited state of the quantum system are, respectively, given by
\begin{subequations}
\begin{eqnarray}
E_{gs}&=&\frac{\pi^2}{6}\frac{\hbar^2}{m w^2}N^3 =
\frac{\pi^2}{6}\frac{L}{w^2\beta^2 A}N^3, \\
\Delta E&=&\pi^2 \frac{\hbar^2}{m w^2}N^{\phantom\dagger}=
\pi^2 \frac{L}{w^2 \beta^2 A}N
\end{eqnarray}
\end{subequations}
\noindent for $N \gg 1$. The partition function of a long system is dominated by the term $\exp \left(-E_{gs} \mathcal{T}/\hbar\right)$.

Thus, the free energy density of a long system of noncrossing polymers is given, to leading order, by
\begin{eqnarray} \label{eq:freeenergy}
\frac{\mathcal{F}}{w L}=\frac{\pi^2}{6} \left( \frac{N}{w} \right)^3 \frac{1}{\beta^2 A}.
\end{eqnarray}
It is straightforward to show that, in contrast, 
the free energy density of a system of noninteracting polymers would be smaller by a factor of $3/N^2$.
The noncrossing conditions essentially restricts each polymer to a region of width of order $w/N$, leading to a strong reduction in entropy and therefore increase in free energy.  Even a weak inter-polymer repulsion would suffice to generate a free energy proportional to $N^3$ rather than $N$.

The ground state energy is proportionate to the mean square polymer slope $\langle \left(\partial_\tau x_n\right)^2\rangle$. This term must be small in order to justify our expression for the deflection energy which necessitates 

\begin{eqnarray} \label{eq:bendingcondition}
\frac{A}{T} \gg \frac{N}{w}.
\end{eqnarray}

\noindent We conclude this section on ground state dominance by remarking that
the lengthscale over which the ground-state dominance approximation holds is $\tau \gg L/\beta \Delta E$, or
\begin{eqnarray} \label{eq:gsdlengthscale}
\tau \gg 
\frac{w^2 \beta A}{N}.
\end{eqnarray}

\section{Topologically constrained systems of polymers}
\label{sec:toposection}

    \subsection{Effects of a single pin on noncrossing polymers}

        \subsubsection{Constraints on the partition function due to a single pin}

\begin{figure}[hh]
\centerline{
\includegraphics[width=.48\textwidth]{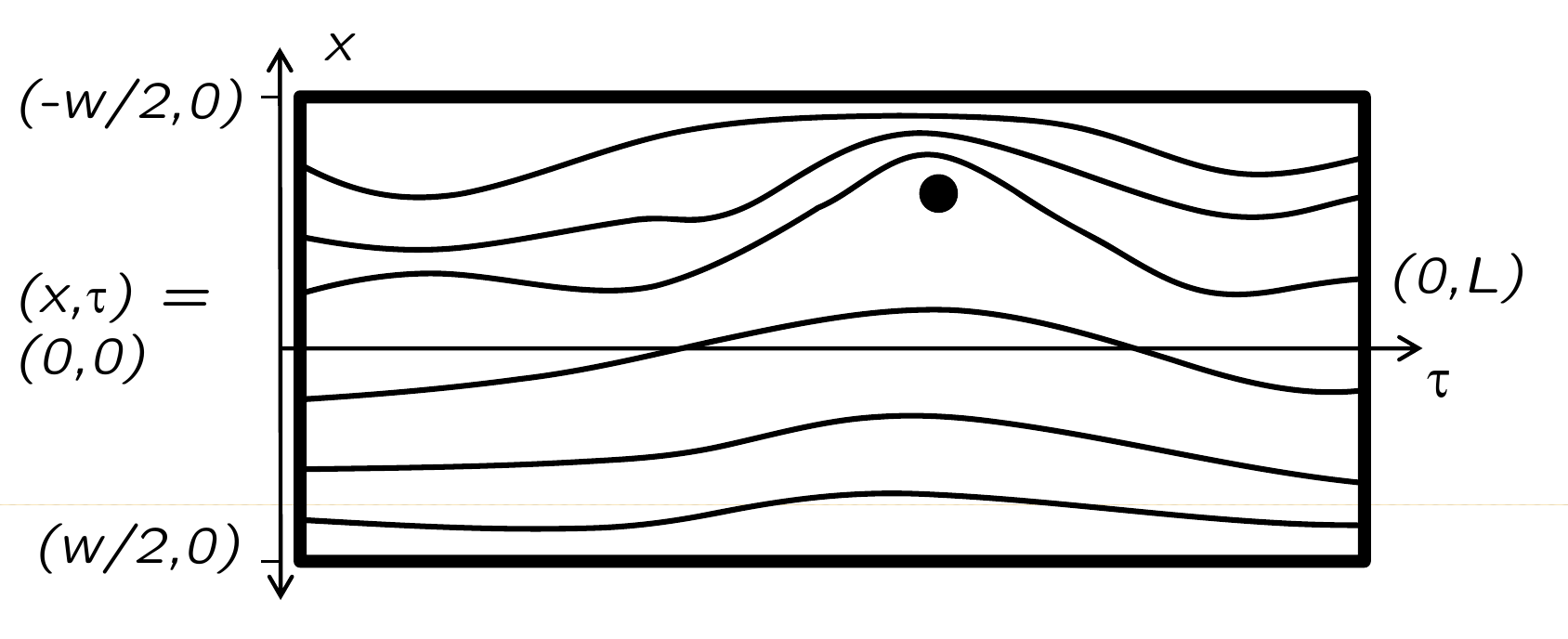}}
\caption{
A topological obstruction (a \emph{pin}), located at $(x,\tau)=(x_p,\tau_p)$. Thermal fluctuations cannot carry a polymer across this pin, so a fixed number of polymers $N_L$ pass to one side of the pin, and the remainder $N_R\equiv N-N_L$ pass to the other side.
}
\label{fig:pindiagram}
\end{figure}

So far, we have been considering systems of polymers that are, from the geometrical and topological standpoints,  trivial.  We now turn our attention to systems that are subject to a \emph{pin}; as we shall see, this renders them nontrivial, topologically.  By a pin we mean a region of the polymer system, sharply localized near the point $(x,\tau)=(x_{p},\tau_{p})$, at which the one-body potential experienced by any polymer segment is taken to be so large and repulsive that the polymers never cross it during the course of an experiment.  Thus, the pin serves as a {\it topological constraint\/} and, because the polymers are directed, {\it it partitions the configuration space\/} of the polymer system into sectors labeled by the number $N_{L}$ of polymers that have the property that as they pass through the line $\tau=\tau_{p}$ they obey $-\pwidth/2< x(\tau_{p})<x_{p}$; see Fig.~\ref{fig:pindiagram}.  Then, the corresponding number of polymers that pass the pin on its other side [i.e., obey $x_{p}<x(\tau_{p})<\pwidth/2$] is given by $N_{R}\equiv N-N_{L}$.  We note that on the line $\tau=\tau_p$ the mean polymer densities to the left and right of the pin are, respectively, $\rho_L = N_L/\big((w/2)+x_p\big)$ and $\rho_R = N_R/\big((w/2)-x_p\big)$.  Evidently, the constraint created by the pin eliminates polymer configurations from the thermal ensemble, and thereby reduces the entropy of the system.  For a generically located pin, the only configurations remaining in the ensemble correspond to large fluctuations of the original system.  The pin therefore raises the free energy of the system and, as a result, there is generically an equilibrium force on the pin.

To determine the increase in the free energy due to the presence of the pin, it is convenient to analyze the partition function of the polymer system, restricted to having $N_{L}$ polymers constrained to pass on one side of the pin (as described more precisely in the previous paragraph), normalized by the unrestricted partition function.  This amounts to computing Eq.~(\ref{eq:observable}) with the observable
$O[\{x_n(\cdot)\}]$
given by
$\delta
\Big(
N_{L},
\sum_{n=1}^{N}
\theta\big(x_{p}-x_{n}(\tau_{p})
\Big)$,
where $\delta(N,N^{\prime})$ is the Kronecker delta function (i.e., $1$ for $N=N^{\prime}$ and $0$ for $N\ne N^{\prime}$) and $\theta(\cdot)$ is the usual Heaviside step function, which takes the value $0$ and $1$, respectively, for negative and positive arguments.  We now set about computing this free energy increase, as well as the impact of the pin on the spatial variation of the polymer density.

\subsubsection{How the constraint is reflected in the quantum propagator}

To compute various consequences of the presence of the pin we employ the mapping from the polymers to  quantum-mechanical particles.  This necessitates that we focus on the propagator, interrupted at imaginary time $\qtime_{p}$ at which time we require that there be precisely $N_{L}$ particles in the range $-\pwidth/2<x<x_{p}$ but with their exact locations left otherwise unspecified, i.e.,
\begin{eqnarray}
&&
\int_{\cal C}
d\{x_{n}\}\,
\langle\Psi_{f}\vert
{\rm e}^{-\qham(\qTime-\qtime_{p})/\hbar}
\vert\{x_{n}\}\rangle
\nonumber\\
&&\qquad\qquad\qquad\qquad\times
\langle\{x_{n}\}\vert
{\rm e}^{-\qham\qtime_{p}/\hbar}
\vert\Psi_{i}\rangle.
\label{eq:interrupt}
\end{eqnarray}
Here, the symbol ${\cal C}$ indicates the following constraints on the range of integration:
\begin{eqnarray}
&&-\pwidth/2<x_{1},\ldots,x_{N_{L}}<x_{p}
\nonumber\\
&&\qquad\qquad\qquad
<x_{N_{L}+1},\ldots,x_{N}<\pwidth/2.
\end{eqnarray}
In quantum-mechanical language, we are therefore to compute two factors: the {\it amplitude\/} to have precisely $N_{L}$ particles at specific positions to the left of the pin (and the complement of particles at specific positions to the right of the pin) at imaginary time $\qtime_{p}$, given that the system was in the state $\vert\Psi_{i}\rangle$ at imaginary time $0$; and the {\it amplitude\/} to have the state $\vert\Psi_{f}\rangle$ at imaginary time $\qTime-\qtime_{p}$, given that precisely $N_{L}$ particles were at specific positions to the left of the pin (and the complement of particles at specific positions to the right of the pin) at imaginary time $0$.

Next, we invoke ground-state dominance, discussed in Sec.~\ref{sec:eigfcn}, which is justified by the restriction to values of the time $t_{p}$ that are far from the initial and final times (i.e., $0$ and $\qTime$).  Thus, the amplitude of Eq.~(\ref{eq:interrupt}), now normalized so as to yield the quotient of partition functions discussed in Sec.~\ref{sec:partitionfcn}, becomes
\begin{eqnarray} \label{eq:constrainedintegration}
&&\int_{\cal C}d\{x_{n}\}\,
\langle\Psi_{gs}\vert\{x_{n}\}\rangle\,
\langle\{x_{n}\}\vert\Psi_{gs}\rangle
\nonumber\\
&&\qquad\qquad\qquad
=\int_{\cal C}d\{x_{n}\}\,
\big\vert\langle\{x_{n}\}\vert\Psi_{gs}\rangle\big\vert^{2};
\label{eq:interGSD}
\end{eqnarray}
the exponential factor $\exp\big({-E_{gs}\qTime/\hbar}\big)$ has canceled between the numerator and denominator, and $\vert\Psi_{gs}\rangle$ is the normalized ground state of the quantum system, the normalization being given by
$\langle\Psi_{gs}\vert\Psi_{gs}\rangle=1$.

\begin{widetext}

\subsubsection{Transformation to the density profile}
\label{sec:transformationdensity}

To perform the constrained integration we invoke the form of the ground-state wave function, Eq.~(\ref{eq:gswavefcn}), and thus rewrite Eq.~(\ref{eq:constrainedintegration}) as (choosing now units so that $\pwidth = \pi$)
\begin{eqnarray} \label{eq:discretefe}
&&\int_{\cal C}d\{x_{n}\}\,
\vert
\langle\{x_{n}\}\vert\Psi_{gs}\rangle\vert^{2}
=
\frac{2^{N^2}}{\pi^{N} N!}
\int_{\cal C}d\{x_{n}\}\,
\exp
\bigg(
\sum_{n=1}^{N}\ln\cos^{2}x_{n}
+
\sum_{1\le n<n^{\prime}\le N}\ln
    \Big[\big(\sin x_{n}-\sin x_{n^{\prime}}\big)^{2}\Big]
\bigg).
\label{eq:interWFexp}
\end{eqnarray}
This (generically high-dimensional) integral is conveniently analyzed by exchanging the discrete particle coordinates $\{x_{n}\}$ for the continuum density function $\pdensity$, via
\begin{eqnarray} \label{eq:rhodef}
\pdensity(\{x_{n}\})=\frac{1}{N}\sum_{n=1}^{N}\delta(x-x_{n}),
\end{eqnarray}
and replacing the multiple integration over polymer coordinates by functional integration over the configurations of the polymer density, via which we obtain 
\begin{eqnarray} \label{eq:rhoeq}
&&\int_{\cal C}
\!\!
d\{x_{n}\}\,
\vert
\langle\{x_{n}\}\vert\Psi_{gs}\rangle\vert^{2}
\!=\!
\frac{2^{N^2}}{\pi^{N}N!}\!
\int_{\widetilde{\cal C}}{\cal D}\pdensity\,
\exp\!
\bigg(\! N
\!\!\!\!
\int\limits_{-\pi/2}^{\pi/2}
\!\!\!
dx\,\pdensity(x)\ln\cos^{2}x
+\!
\frac{N^{2}}{2}
\!\!\!
\int\limits_{-\pi/2}^{\pi/2}
\!\!\!\!\!
dx\,dx^{\prime}
\pdensity(x)\,
\pdensity(x^{\prime})
\ln \Big[ \big(\sin x-\sin x^{\prime}\big)^{2} \Big]
\!\bigg)\!. \quad
\label{eq:interWFcts}
\end{eqnarray}
Here, the symbol $\widetilde{\cal C}$ indicates the condition on $\pdensity$ that corresponds to the pin constraint ${\cal C}$, viz.,
\begin{subequations}
\begin{eqnarray}
\int_{-\pi/2}^{x_{p}}dx\,\pdensity(x)=N_{L}/N,
\label{eq:pincon}
\end{eqnarray}
but also the
normalization and positivity conditions, which are respectively given by
\begin{eqnarray}
\int_{-\pi/2}^{\pi/2}dx\,\pdensity(x)&=&1;
\label{eq:unitnorm}
\\
\pdensity(x)&\ge&0,\quad\mbox{for}\quad-\pi/2<x<\pi/2;
\label{eq:positivity}
\end{eqnarray}
\end{subequations}
both of which follow from the definition of $\pdensity$ as a density, Eq.~(\ref{eq:rhodef}).  Thus, in Eq.~(\ref{eq:rhoeq}) we have arrived at a convenient formulation of the partition function in the presence of the pin constraint.  The convenience results from the replacement of particle coordinates by the collective density field, which enables us to employ the techniques of the calculus of variations, and thus to determine the value of the density that yields the dominant (in the sense of Laplace's method) contribution to the functional integral.  A more rigorous transition from individual polymer coordinates to a density function via the method of collective coordinates~\cite{doiedwards} would yield an additional $O(N\ln N)$ term in the exponential in Eq.~(\ref{eq:rhoeq}).  Our current level of accuracy suffices when all length scales, such as the displacement of the pin from its equilibrium position, are long compared to the average inter-polymer spacing.

It should be noted that in Eq.~(\ref{eq:rhoeq}) we have an effective long-ranged (logarithmic) repulsion between polymer segments, which replaces the original, local interaction.  Two polymers that are laterally separated at $\tau=\tau_p$ will nevertheless interact indirectly with one another, owing to the nonzero probability that they would encounter one another at some distant value of $\tau$.  Thus, as we are effectively integrating out all polymer degrees of freedom that do not lie on the line $\tau=\tau_p$, we are generating an additional inter-polymer repulsion.  We have thus mapped our two-dimensional system with its short-ranged interactions on to a one-dimensional system having long-ranged interactions.  
Such a description is adequate for addressing large fluctuation phenomena on the line $\tau=\tau_p$
that arise from a pin or other constraints.

\end{widetext}

\subsubsection{Obtaining the density profile}
\label{sec:densityprofile}

The factors of $N$ in the exponent of Eq.~(\ref{eq:rhoeq}) suggest that we may treat this functional integral via the functional version of Laplace's method, and therefore via finding the maximum of the leading term of the exponent.  To this end, we approximate the partition function as
\begin{eqnarray}
\mathcal{Z} \sim \max_{\widetilde{ \cal C}} \exp \left(- \frac{\Delta\mathcal{F}\left[\pdensity\right]}{T}\right)
\end{eqnarray}
where the free energy functional describing the effect of the pin
$\Delta\mathcal{F}\left[\pdensity\right]$
is defined via collecting the leading terms in Eq.~(\ref{eq:rhoeq})

\begin{eqnarray}
\label{eq:deltaf}
  \Delta\mathcal{F}\!\left[\pdensity\right]
\!\equiv\!
-N^2 T \ln 2 \qquad \qquad \qquad \qquad \qquad \qquad  \\
\!\!\!\!\!
- \frac{N^2 T}{2} \int\limits_{-\pi/2}^{\pi/2}
\!\!\!\!
dxd\dummyxprime
\pdensity(x)
\pdensity(\dummyxprime)
\ln\!\Big[\left(\sin x\!-\!\sin\dummyxprime\right)^{2}\Big]\!. \nonumber
\end{eqnarray}

We note that this expression for the free energy scales with $N$ as $N^2$.  As they do for the overall free energy of the system, the inter-polymer interactions qualitatively change the response of the polymers to the pin.

Our next task is determine the approximate value of the mean density profile $\mdensity(x)$ that minimizes the free energy, subject to the constraints discussed in Sec.~\ref{sec:transformationdensity}.  Upon implementing the constraints~(\ref{eq:pincon},\ref{eq:unitnorm}) via Lagrange's method of undetermined multipliers, we arrive at the following first-order stationarity condition:
\begin{eqnarray}
\label{eq:stationarity}
&&\int_{-\pi/2}^{\pi/2}
dx\,\mdensity(x)\,
\ln\vert\sin x-\sin \dummyxprime\vert
\nonumber\\
&&\qquad\qquad\qquad\qquad
+\lambda_{1}+\lambda_{2}\,\theta(x_{p}-\dummyxprime)=0.
\label{eq:first-order}
\end{eqnarray}
It is important to note that this first-order condition is only required to hold at values of $x$ for which $\mdensity(x)$ {\it is greater than zero\/}, owing to the positivity condition~(\ref{eq:positivity}) that $\pdensity$ is required to obey.  In the absence of the pin constraint, the $\lambda_2$ term would be absent, and the condition would be trivially satisfied via a uniform polymer density profile.
This uniform density profile is also the correct mean density profile in the presence of a pin for the exceptional case in which the pin demands that the average polymer densities on the two sides of it be equal to one another (i.e., for the case in which $\rho_L=\rho_R=N/w$.  We refer to such a constraint as an \emph{equilibrium pin}.  Generically, however, the pin has a profound effect on the mean density profile.

To analyze the stationarity condition, Eq.~(\ref{eq:stationarity}), it is convenient to make the following transformations of the dependent and independent variables:
\begin{subequations} \label{eq:newvar}
\begin{eqnarray}
\mdensity(x)&\rightarrow&
\newpdensity(\newx)\equiv \mdensity(x)/\cos x,\\
x&\rightarrow&
\newx\equiv\sin x,
\end{eqnarray}
\end{subequations}
so that $dx=d\newx/\sqrt{1-\newx^{2}}$,
and similarly for $\dummyxprime$.  In terms of these new variables the stationarity condition becomes
\begin{eqnarray}
&&\int_{-1}^{1}
d\newx\,\newpdensity(\newx)\,
\ln\vert{\newx-\newdummyxprime}\vert
\nonumber\\
&&\qquad\qquad\qquad
+\lambda_{1}+\lambda_{2}\,\theta(\newx_{p}-\newdummyxprime)=0,
\label{eq:new-first-order}
\end{eqnarray}
where $\newx_{p}\equiv\sin x_{p}$.  This form of the stationarity condition must be met at values of $\newx$ for which $\newpdensity(\newx)>0$.

If we were to ignore the positivity condition that a physically acceptable density profile $\newpdensity$ must satisfy  then it would be straightforward to solve the integral equation~(\ref{eq:new-first-order}) for all $\newx$ (by following a technique that we shall, in fact, eventually adopt; see, e.g., Ref.~\cite{estrada}).  However, the result we would obtain for $\newpdensity$ would be invalid, as it would diverge to negative infinity near the pin.  Thus, we search for a solution that violates the first-order conditions over one or more segments of $-1<\newx<1$; within these segments $\newpdensity(\newx)=0$ and we refer to any such a segment as a {\it gap\/}.

Nonequilibrium pins necessarily cause compression of the the polymers either on one side of the pin or on the other.  Without loss of generality, we may take the compressed region to correspond to $-1<\newx<\newx_{p}$; then the rarified region corresponds to $\newx_{p}<\newx<1$.  For the profile that minimizes the free-energy functional~(\ref{eq:deltaf}), it is physically reasonable to assert that any gap would form on the rarified side and, furthermore, that it would lie directly adjacent to the pin and extend over a region that more strongly disfavors the presence of polymers, as a result of its proximity to the dense region of polymers that is created by the pin on the other side of the pin.  (These assertions will be verified as we proceed.)\thinspace\  Thus, we look for non-negative profiles for which we require $\newpdensity(\newx)=0$ for $\newx$ lying within the range $\newx_{p}<\newx<\newx_{g}$.  If a gap were to be present anywhere else, the system free energy could be lowered by moving polymer density into that gap.  Thus, any solution found that (i)~has this gap and (ii)~elsewhere satisfies the constraints on $\mdensity(x)$ is the unique minimizer of our effective free energy.

\begin{figure}[hh]
\centerline{
\includegraphics[width=.48\textwidth]{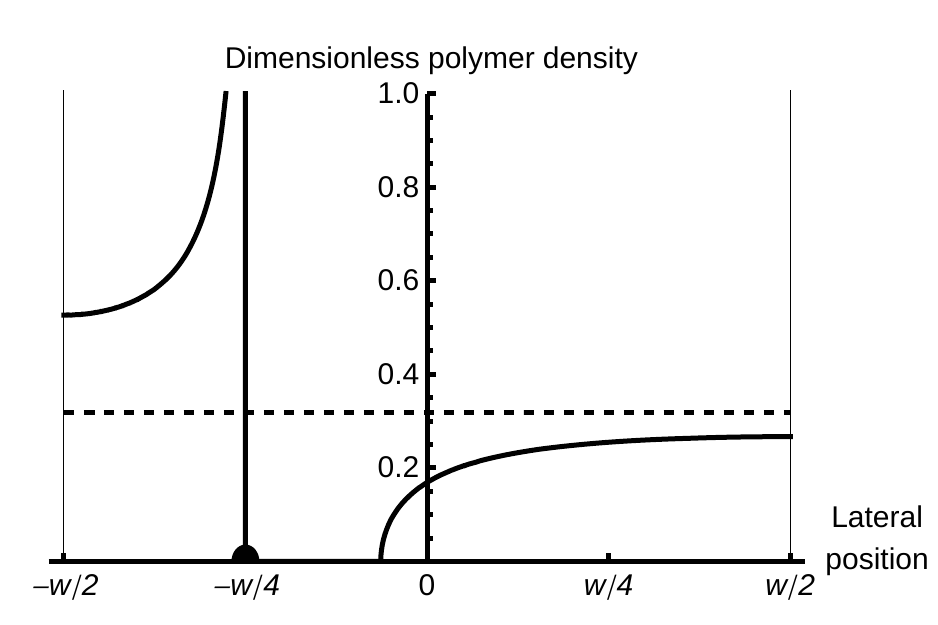}}
\caption{
Transverse variation of the equilibrium directed polymer density along the line $\tau=\tau_p$ for a system subject to a pinning constraint located at $(x,\tau)=(-w/4,\tau_p)$, the location of which is marked by a black dot. 
The physical polymer density is the plotted quantity scaled by $N \pi/ \pwidth$.
Within the present approximation scheme, a finite gap (i.e., an area of zero polymer density) extends from the pin. The dashed line represents the equilibrium polymer density in the absence of a pin.
}
\label{fig:pinplot}
\end{figure}

As we shall see, the formation of a {\it single\/} gap yields a suitable profile (i.e., one that satisfies the positivity constraint automatically).  What remains, then, for the case of a single pin, is to determine the value of the density profile and, in particular, the width of the gap $(\newx_{g}-\newx_{p})$ that together ensure that the necessary conditions on the profile are obeyed by it.  Following the approach reviewed in Ref.~\cite{estrada},
we have that the family of solutions to this integral equation is given in terms of parameters $A_0$ and $A_1$ by
\begin{eqnarray}
\nonumber
\!\!\!\!\!\!
\newpdensity(\newx)\!=\!
\begin{cases}
\!\displaystyle\frac{(A_{0}+A_{1}\newx)\sgn (\newx - \newx_g)}
{\sqrt{(1-\newx^{2})(\newx-\newx_{p})(\newx-\newx_{g})}},
&\begin{matrix}
\!\mbox{for }-1<\newx<\newx_{p}\hfill\\
\!\mbox{or }\newx_{g}<\newx<1;\hfill
\end{matrix}
\\
\noalign{\medskip}
0,
&\!\mbox{for }\newx_{p}<\newx<\newx_{g}.
\end{cases}
\end{eqnarray}
Transforming back to polymer coordinates, using Eq.~(\ref{eq:newvar}), we thus have
$\mdensity(x)$ given by
\begin{eqnarray}
\nonumber
\!\!\!
\begin{cases}
\displaystyle
\frac{(A_{0}+A_{1}\sin x)\sgn(\sin x - \sin x_g)}
{\sqrt{(\sin x-\sin x_{p})(\sin x-\sin x_{g})}},
&\begin{matrix}
\!\!\mbox{for }-\pi/2<x<x_{p}\hfill\\
\!\!\mbox{or }x_{g}<x<\pi/2;\hfill
\end{matrix}
\\
\noalign{\medskip}
0,
&\!\!\mbox{for }x_{p}<x<x_{g}.
\end{cases}
\end{eqnarray}
The next step in determining $\mdensity(x)$ is to adjust $A_{0}$ and $A_{1}$ to ensure that there is no divergence at $x=x_{g}$; this requires that $A_{0}+A_{1}\sin x_{g}=0$.  Physically, this choice is motivated by the expectation that the equilibrium density does not diverge on the rarified side of the pin.  Next, we invoke the normalization condition~(\ref{eq:unitnorm}) and thus determine that $A_{1}=1/\pi$; and, finally, we adjust $x_{g}$ to ensure that the pin constraint~(\ref{eq:pincon}) is met.  Thus, we arrive at the profile that dominates the constrained partition function:
\begin{eqnarray}
\nonumber
\!\!\!
\mdensity(x)=
\begin{cases}
\displaystyle
\frac{1}{\pi}
\sqrt{\frac{\sin x-\sin x_{g}}
{\sin x-\sin x_{p}}},&
\begin{matrix}
\!\mbox{for }-\pi/2<x<x_{p}\hfill\\
\!\mbox{or }x_{g}<x<\pi/2;\hfill
\end{matrix}
\\
\noalign{\medskip}
0,&
\!\mbox{for }x_{p}<x<x_{g}.
\end{cases}
\end{eqnarray}
Restoring the physical lengths, this becomes $\mdensity(x)=$
\begin{eqnarray}
\begin{cases}
\nonumber
\displaystyle
\frac{1}{\pwidth}
\sqrt{\frac{\sin(\pi x/\pwidth)-\sin(\pi x_{g}/\pwidth)}
{\sin(\pi x/\pwidth)-\sin(\pi x_{p}/\pwidth)}},&
\begin{matrix}
\!\mbox{for }-\pwidth/2<x<x_{p}\hfill\\
\!\mbox{or }x_{g}<x<\pwidth/2;\hfill
\end{matrix}
\\
\noalign{\medskip}
0,&
\!\mbox{for }x_{p}<x<x_{g}.
\end{cases}
\end{eqnarray}
Note the essential features of this solution for the polymer density along the line $\tau=\tau_{p}$, as shown in Fig.~\ref{fig:pinplot}: throughout the width of the system (i.e., for $-\pwidth/2<x<\pwidth$/2) the density is non-negative; at the system edges (i.e., $x=\pm \pwidth/2$) the density is finite; at the pin (i.e., for $x=x_{p}$) the density has a square-root divergence approaching $x_p$ from the compressed side and a square-root vanishing approaching $x_g$ from the rarified side; and within the segment $x_{p}<x<x_{g}$ the density is zero.  It is striking that merely as a result of inter-polymer interactions that are local in the two-dimensional plane the topological restriction presented by the pin causes the opening up of a finite gap in the polymer density and, in particular, that the reach of its impact extends over many times the intrinsic inter-polymer separation, at least at the level of the present mean-field type of approximation.  The mechanism responsible for this is that---over a  longitudinal distance that is nonzero---the polymers are energetically disfavored from entirely filling in the gap by the cost in bending energy they would have to incur to depart from their long-distance equilibrium positions.  Note that while the mean density $\mdensity(x)$, as calculated within our approach, is zero within the gap, fluctuations can only increase it, leading to there being a small positive polymer density within this region.

\begin{figure}[hh]
\centerline{
\includegraphics[width=.48\textwidth]{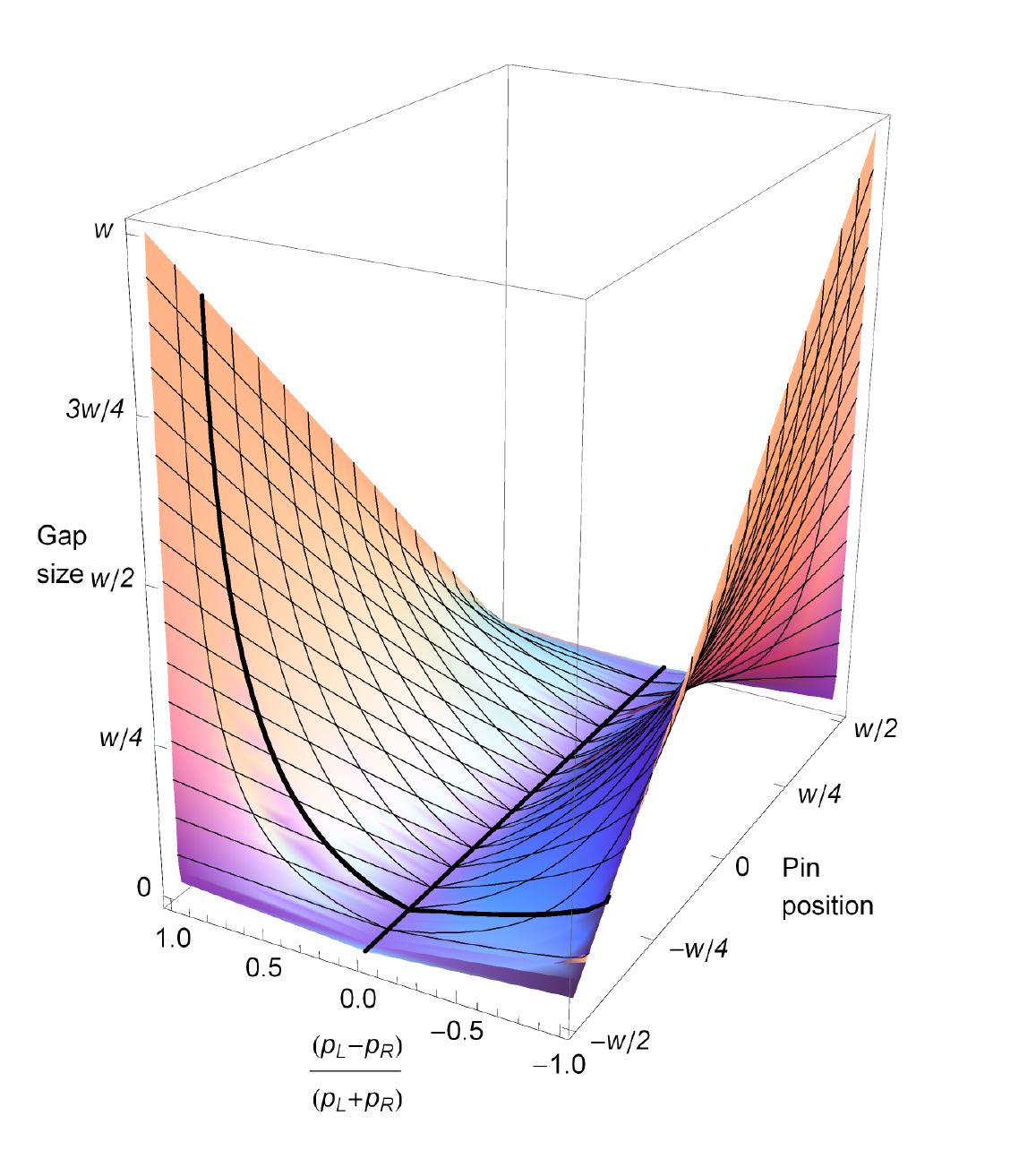}}
\caption{
Dependence of the width of the gap that opens in the polymer density adjacent to a pin on the pin location and polymer density imbalance. Note that the gap width increases from zero as the pin position is varied, parametrically, from its equilibrium position. At the equilibrium position ($\rho_L = \rho_R$), the gap vanishes and the polymer density is spatially uniform, as it is for a system in the absence of a pin.
}
\label{fig:gapfunction}
\end{figure}

A closed-form expression for the gap edge $x_g$ associated with a partitioning may be found in terms of elliptic integrals, as in Eq.~(\ref{eq:sgspnrelationship}). For a pin near its equilibrium position $x_e$, the pin displacement $x_p-x_e \equiv \disp$ and the gap $x_p-x_g\equiv \gap$ are both small. Then the gap size is related to pin displacement to leading order by

\begin{eqnarray} \label{eq:smallgap}
\disp \approx  \frac{\gap}{2} \ln \left( \frac{2 \pwidth}{\pi |\gap|}\cos^2 \frac{\pi x_p}{\pwidth} \right) + O(\gap).
\end{eqnarray}

\noindent Thus for a pin near its equilibrium position the gap size is sublinear in the pin displacement.

In Fig.~\ref{fig:gapfunction} we show the gap size more generally as a function of the pin position and the parameter $\nu\equiv(\rho_L-\rho_R)/(\rho_L+\rho_R)$, which we introduce to characterize the imbalance between the densities on either side of the pin.  
When all the polymers are on one side of the pin (in which case $\nu=\pm1$) then the gap size is simply the pin displacement.  More generally, the polymers on the rarefied side expand towards the pin so that the gap is smaller than the pin displacement.

\subsection{Force on a pin}

\subsubsection{Calculation of the force}

Having determined the mean density profile on the lateral line through the pin, viz.~$\pdensity(x,\tau)\vert_{\tau=\tau_{p}}$, we now return to our formulation of the dominant contribution to the partition function, and hence the free energy, to determine the increase in the free energy due to the pin, viz.~$\Delta\mathcal{F}$.  To do this, we begin with the free energy functional Eq.~(\ref{eq:rhoeq}) and seek to compute its value at the mean density profile.  To simplify this computation, we employ the first-order stationarity condition~(\ref{eq:stationarity}) to eliminate the combination
$-\int_{-\pwidth/2}^{\pwidth/2}
dx^{\prime}\,\pdensity(x^{\prime})\,
\ln\big\vert{\sin (\pi x/\pwidth)-\sin (\pi x^{\prime}/\pwidth)}\vert$
in favor of
$\lambda_{1}+\lambda_{2}\,\theta(x_{p}-x)$.
We also use the stationarity condition evaluated at $x=\pm\pwidth/2$ to obtain the Lagrange multipliers $\lambda_{1}$ and $\lambda_{2}$ in terms of the (known) mean profile.  Then, in the resulting expression for $\Delta\mathcal{F}$, we use the normalization of $\pdensity$, Eq.~(\ref{eq:unitnorm}), and the constraint on it that the pin introduces, Eq.~(\ref{eq:pincon}), and thus arrive at the result
\begin{eqnarray}
&&\Delta\mathcal{F}(x_{p},N_{L},N,T,\pwidth)
=
\nonumber\\
&&\quad
-N^{2}T\int_{-\pwidth/2}^{\pwidth/2}dx\,\mdensity(x,x_{p},x_{g})\,
\nonumber\\
&&\qquad\quad
\times
\Big(
 (N_{L}/N)\ln 2 \big(1+\sin(\pi x/\pwidth)\big)
\nonumber\\
&&
\qquad\qquad\quad
+(N_{R}/N)\ln 2 \big(1-\sin(\pi x/\pwidth)\big)
\Big),
\label{eq:FEincrease}
\end{eqnarray}
\noindent 
We remind the reader that the gap edge location $x_{g}$ is not an independent variable but is determined in terms of the independent variables, via the pin constraint, Eq.~(\ref{eq:pincon}).

As a special case, we first consider the situation in which all polymers lie to one side of the pin, i.e.~$N_{L}=N$, so that the pin can in effect be taken to be a septum emerging normally from one wall of the system.  In this case, the formula for the free energy increase, Eq.~(\ref{eq:FEincrease}), simplifies to the explicit form
\begin{eqnarray} \label{eq:nreqz}
&&\Delta\mathcal{F}=
-\frac{N^{2}T}{w}\int_{-\pwidth/2}^{x_{p}}dx\,
\sqrt{\frac{\sin(\pi x/\pwidth)-1}
{\sin(\pi x/\pwidth)-\sin(\pi x_{p}/\pwidth)}}
\nonumber\\
&&
\qquad\qquad\qquad
\qquad\qquad\qquad
\times\ln 2 \big(1+\sin(\pi x/\pwidth)\big)
\nonumber\\
\noalign{\smallskip}
&&\qquad\qquad
=-N^2 T \ln\big(\left[1+\sin(\pi x_{p}/\pwidth)\right]/2\big).
\end{eqnarray}
Remaining with the case $N_{L}=N$, we note that for mild compressions (i.e, those obeying $(\pwidth/2)-x_{p}\ll\pwidth$) the force exerted on the pin is Hookean in nature:
\begin{eqnarray}
\Delta\mathcal{F}=
\frac{\pi^{2}}{4}N^2 T
\left(\frac{(\pwidth/2)-x_{p}}{\pwidth}\right)^{2},
\end{eqnarray}
as established by making Taylor expansion of Eq.~(\ref{eq:nreqz}) about $x=\pwidth/2$.
On the other hand, for strong compressions [for which $(\pwidth/2)+x_{p}\ll\pwidth$], so that the polymers are forced through an opening the width of which is only a small fraction of the full width of the system, we have the form
\begin{eqnarray}
\Delta\mathcal{F}
=2 N^{2}T\,
\ln\left(
\frac{2}{\pi}
\left(
\frac{\pwidth}{(\pwidth/2)+x_{p}}
\right)
\right).
\end{eqnarray}
Residing, as it does, beyond the linear-response regime, it is not surprising that this form is non-Hookean. Indeed, a similar term, with $N_L^2$ replacing $N^2$, dominates the free energy for highly compressed systems with polymers on both sides of the pin.

\begin{figure}[hh]
\centerline{
\includegraphics[width=.48\textwidth]{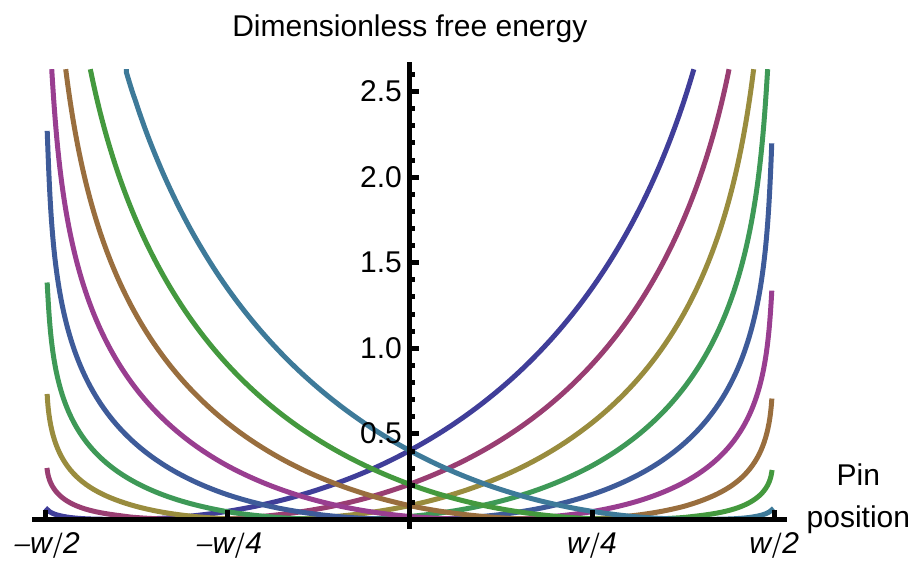}}
\caption{\label{phase}
Increase in free energy as a function of pin position, for various values of the number of directed polymers $N_L$ passing to one side of the pin.  The physical free energy is the plotted quantity scaled by $N^2 T$. The free energy cost due to the pin is less than harmonic for a generic pin that is near its equilibrium position, and it diverges logarithmically as the pin nears an edge of the system.  The curves plotted here are for pins having equilibrium (and free energy-minimizing) positions
$-.4 \pwidth, \ -.3 \pwidth, \ -.2 \pwidth \ \ldots \ .4 \pwidth$.
}
\label{fig:feplot}
\end{figure}

In the more general case, in which $N_R>0$ polymers pass to the right of the pin, the force on the pin
obeys (see appendix~\ref{sec:ellipticals})

\begin{eqnarray} \label{eq:fedifeq}
-\frac{d \mathcal{F}}{d x_p}=-\frac{\pi N^2 T}{\pwidth} \left(\frac{\sin \left( \pi x_p/\pwidth\right)- \sin \left(\pi x_g/\pwidth\right)}{\cos{\left(\pi x_p/\pwidth\right)}}\right)\!. \qquad
\end{eqnarray}

\noindent For a pin near its equilibrium position this force is simply proportionate to the gap size $x_p-x_g$, and the above expression may be integrated to obtain the leading term in the free energy

\begin{eqnarray} \label{eq:feexpansion}
\mathcal{F} \approx \frac{N^2 T}{4} \left(\frac{\pi \gap}{\pwidth}\right)^2 \ln \left[ \frac{2 \pwidth}{\pi |\gap|} \cos^2 \frac{\pi x_p}{\pwidth}\right]\! . \quad
\end{eqnarray}

\noindent Note that although the free energy of the pin near its equilibrium position grows faster than quadratically in the gap size $\gap$ it is sub-Hookean in (i.e., grows slower than quadratically with)the pin displacement $x_p-x_e$. In contrast, in the limit in which the pin position approaches the boundary of the system and highly compresses the polymers, the free energy diverges logarithmically.
Fig.~\ref{fig:feplot} shows the free energy as a function of pin position for polymer partitionings corresponding to various equilibrium positions of the pin.

\subsection{Effects of a barrier on noncrossing polymers}

\begin{figure}[hh]
\centerline{
\includegraphics[width=.48\textwidth]{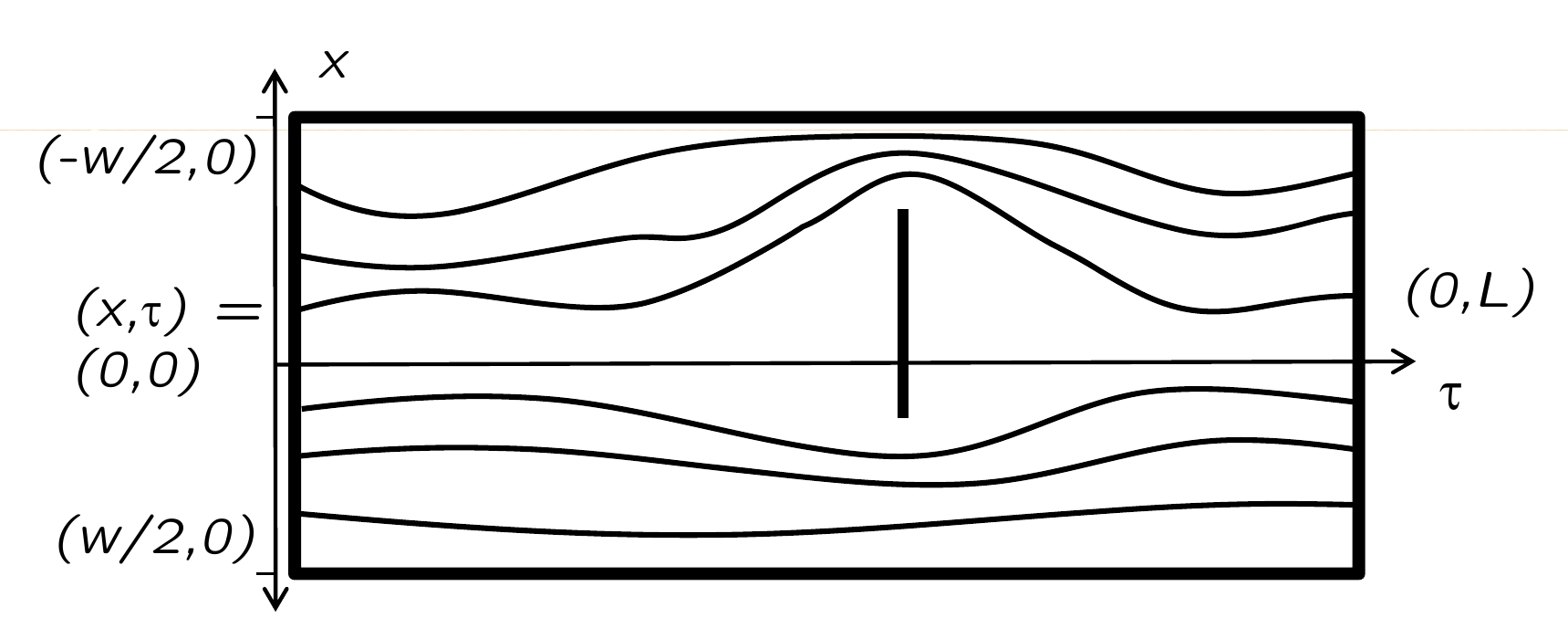}}
\caption{\label{phase}
A barrier (i.e., a laterally oriented topological obstruction of nonzero width) that constrains the number of directed polymers that pass on either side.  By using the techniques developed in the present paper one can readily analyze the effects of such barriers.
}
\label{fig:barrierdiagram}
\end{figure}

The approach we have developed so far addresses the case of a topological constraint created by an infinitesimally wide pin.  We now generalize the approach to allow consideration of the case in which the constraint is a barrier of finite width, so that polymers are prohibited from passing through an extended line segment
$(x,\tau)\vert_{\tau=\tau_{p}}$ (with $-\pwidth/2\le x_{p}^{L}<x<x_{p}^{R}\le\pwidth/2$), as shown in Fig.~\ref{fig:barrierdiagram}.
We note that the effect of single barrier is, from the standpoint of directed-polymer statistical mechanics, entirely equivalent to the effect of a pair of pins, provided that polymers are forbidden from passing between the pins.

To analyze the situation of a barrier and, in particular, to compute the change in free energy arising from the presence of this barrier, we return to the task of minimizing the free energy~(\ref{eq:deltaf}), but we replace the pin constraints~(\ref{eq:pincon},\ref{eq:unitnorm}), by the following ones, appropriate for a barrier in which precisely $N_{L}$ polymers pass to the left of the barrier and $N_{R}$ pass to its right:
\begin{subequations}
\begin{eqnarray}
\int_{-\pi/2}^{x_{p}^{L}}  dx\,\pdensity(x)&=&N_{L}/N,\\
\int_{x_{p}^{R}}^{\pi/2}dx\,\pdensity(x)&=&N_{R}/N.
\end{eqnarray}
\end{subequations}
These constraints along with normalization and positivity imply the barrier condition:
$\int_{x_{p}^{R}}^{x_{p}^{L}}  dx\,\pdensity(x)=0$.

Physically, it is evident that there are two distinct situations.  Consider a barrier of given width, and with a partitioning specified by $N_{L}$ and $N_{R}$.  Relative to the situation without the barrier, the polymers are compressed on at least one side of the barrier, and possibly both.  Let us focus on a compressed side, at which the polymer density diverges, and imagine shrinking the barrier into a pin located at $x_P^L$.  On the other side of the pin, there would now be a gap in the polymer density, the width of which is determined, as before, by the pin constraints.  Now, imagine widening the pin into a barrier.  As long as the barrier width does not exceed the gap width, the polymer density profile would not change, remaining at the density profile associated with a single pin at $x_P^L$.  In effect, the barrier resides in the gap created by the pin, so the fact that the barrier width is finite has no impact.

\begin{figure}[hh]
\centerline{
\includegraphics[width=.48\textwidth]{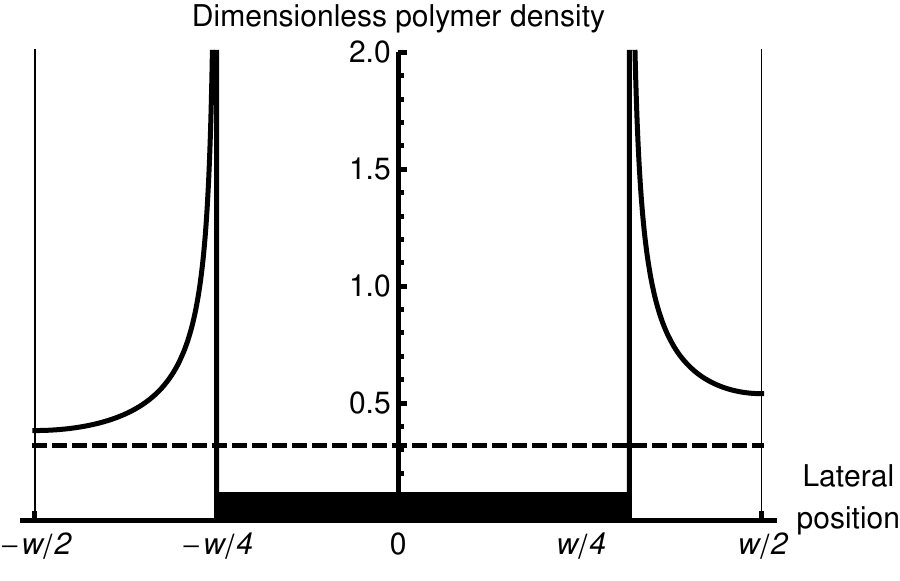}}
\caption{\label{phase}
Equilibrium directed-polymer density along the line containing a barrier.  
The physical polymer density is the plotted quantity scaled by $N \pi/ \pwidth$.
The spatial extent of the barrier is indicated by the thick line.  
Note that for the chosen value of the polymer partitioning, the polymer density diverges on both sides of the barrier.
The dashed line represents the density of the polymer system in the absence of the barrier.
}
\label{fig:barrierplot}
\end{figure}

The second situation follows when the barrier width {\it does\/} exceed the width of the gap created by the pin.  Now, both ends of the barrier are in contact with polymers.  Repeating the integral-equation analysis of Sec.~\ref{sec:densityprofile}, we find the resulting density profile (choosing units as before so that $\pwidth = \pi$) to be given by 
$\mdensity(x)=$
\begin{eqnarray}
\nonumber
\!\!\!
\begin{cases}
\displaystyle
\frac{(A_{0}+A_{1}\sin x)\sgn(\sin x - \sin x_{p}^{R})}
{\sqrt{(\sin x-\sin x_{p}^L)(\sin x-\sin x_{p}^R)}},
&\begin{matrix}
\!\!\mbox{for }-\pi/2<x<x_{p}^L\hfill\\
\!\!\mbox{or }x_{p}^R<x<\pi/2;\hfill
\end{matrix}
\\
\noalign{\medskip}
0,
&\!\!\mbox{for }x_{p}^L<x<x_{p}^R.
\end{cases}
\end{eqnarray}
The two constants are now set by the numbers of polymers passing on either side of the barrier. The area of the \lq\lq gap\rq\rq\ is now precisely the area excluded by the barrier, and the polymer density diverges on either side of it, as shown in Fig.~\ref{fig:barrierplot}.

Although, in general, the effect of the barrier on the free energy is more complicated than the effect of a single pin, there is one case that may be addressed analytically.  Consider an extended barrier of width $b \ll w$, the mid-point of which would be an equilibrium pin if $b$ were set to $0$.  In this case, the equilibrium polymer density on either side of the barrier resembles the polymer density around a single pin that has been displaced a distance $b/2$ from an equilibrium position $x=w/2$. The free energy cost of such a small barrier is then given by
\begin{eqnarray}
\Delta\mathcal{F}=
\frac{\pi^{2}}{8}\,T \left(\frac{N}{w}\right)^{2}b^{2}.
\end{eqnarray}

\subsection{Effects of multiple pins and/or barriers on noncrossing polymers}

We now have all the tools we need to address multiple pins and even multiple finitely-wide barriers, as long as we continue to restrict these obstacles to lying on a common line $\tau=\tau_p$.  Consider, then, $M$ pins at the ordered locations $-(\pwidth/2)<x_{1}<x_{2}<\cdots<x_{M}<(\pwidth/2)$, such that $N_\ell$ polymers are constrained to pass between pin $\ell$ (located at $x_{\ell}$) and the nearest pin (or wall) to its left (i.e., at smaller $x$).  $N_1$ and $N_{M+1}$ respectively denote the number of polymers passing to the left of the leftmost pin and the right of the rightmost pin.
(We remind the reader that any barrier of finite width may be treated as an adjacent pair of pins with no polymers passing between them.)\thinspace\ For such situations, the cost in free energy is determined, as usual, by maximizing the logarithm of the appropriate partition function $\mathcal{Z}$, given by (choosing here units so that $\pwidth = \pi$)
\begin{eqnarray}
&&\ln\mathcal{Z}
\!\sim\!\!
\int_{\cal C}\!\! dx\!\!
\int_{\cal C}\!\! d\bar{x}\,
\rho(x)\,\rho(\bar{x})
\ln \left[\big(\sin x-\sin\bar{x}\big)^2\right]\!, \quad
\end{eqnarray}
over the density profile $\pdensity(\cdot)$, subject to the constraints imposed by the pins and/or barriers.  The symbol ${\cal C}$ now indicates the following collection of constraints for all $\ell$:
\begin{eqnarray}
\int_{{\cal C}_{\ell}}dx\,\pdensity(x)
&=&
N_{\ell}/N,
\\
\pdensity(x) &\geq& 0,
\quad
\mbox{for} \quad -(\pi/2) < x < (\pi/2), \quad
\end{eqnarray}
where ${\cal C}_{\ell}$ indicates that the integration range runs to the $\ell^{\,\rm th}$ obstacle from the obstacle or wall that precedes it (or, in the rightmost case, from the rightmost obstacle to the right wall).
The barrier constraints are implemented via a collection of Lagrange multiplier terms, which augment $\ln{\cal Z}$ and are given by
\begin{eqnarray}
\sum_{\ell=1}^{M+1}
\lambda_{\ell}
\left\{
\int_{{\cal C}_{\ell}}dx\,\pdensity(x)-\frac{N_\ell}{N}
\right\}.
\end{eqnarray}

As with the cases already treated in Sec.~\ref{sec:densityprofile}, functional differentiation---of the free energy functional augmented by the Lagrange multiplier terms---yields an integral equation that is solvable for a non-negative, and therefore physically acceptable, $\pdensity(\cdot)$, provided we (i)~allow for the possibility of gaps in the polymer density profile, and (ii)~determine their necessity and location by implementing the constraints on the numbers of polymers passing between the various obstacles. In general, some pins will lie within the interior of gaps; other pins will have a gap form on one side.

For cases involving more than one or two pins, the process of determining the gap structure that permits all constraints to be satisfied becomes quite tedious, but it should always yield a unique result for the density profile that minimizes the free energy.
Moreover, denoting by $\{{p_j}\}$ the sets of points at which the polymer density diverges
and by $\{g_k\}$ the set where it increases gradually from zero density,
we may continue as before and thus obtain the mean polymer density in regions outside the gaps as
\begin{eqnarray}
\mdensity(x)=
\frac{1}{\pwidth}
\sqrt{ \frac{\prod_k [\sin(\pi x/\pwidth)-\sin(\pi g_k/\pwidth)]}
{ \prod_j [\sin(\pi x/\pwidth)-\sin(\pi p_j/\pwidth)]}}.
\end{eqnarray}

\subsection{Connection to quantum hydrodynamics}

In describing the effect of a single pin, we have focused specifically on the equilibrium structure of the polymer system on the line $\tau=\tau_p$ that passes though the pin.  To describe the polymer behavior away from this line (i.e., for $\tau\ne\tau_p$) is equivalent to describing the imaginary-time evolution of the quantum system away from this line that is consistent with the density profile $\mdensity(x)$ at $\tau=\tau_p$.  As this profile differs substantially from the equilibrium density profile for the pin-less case (i.e., a uniform density profile) the equivalent quantum system can be regarded as having undergone a large quantum fluctuation.

Systems of nonlinear equations analogous to hydrodynamical equations have been used to describe the evolution of one-dimensional systems of interacting particles around large fluctuations~\cite{abanov}.  In terms of the particle density and velocity fields, $\rho(x,t)$ and $v(x,t)$, these equations are
\begin{subequations}
\begin{eqnarray}
\partial_t \rho + \partial_x \left(\rho v \right) &=& 0, \\
\partial_t v + v \partial_x v &=& m^{-1} \partial_x \, \partial_\rho \left( \rho \mathcal{E}(\rho)\right),
\end{eqnarray}
\end{subequations}
where $\mathcal{E}(\rho)$ is the ground-state energy per particle, expressed as a function of the density $\rho$.

\begin{figure}[hh]
\centerline{\includegraphics[width=.48\textwidth]{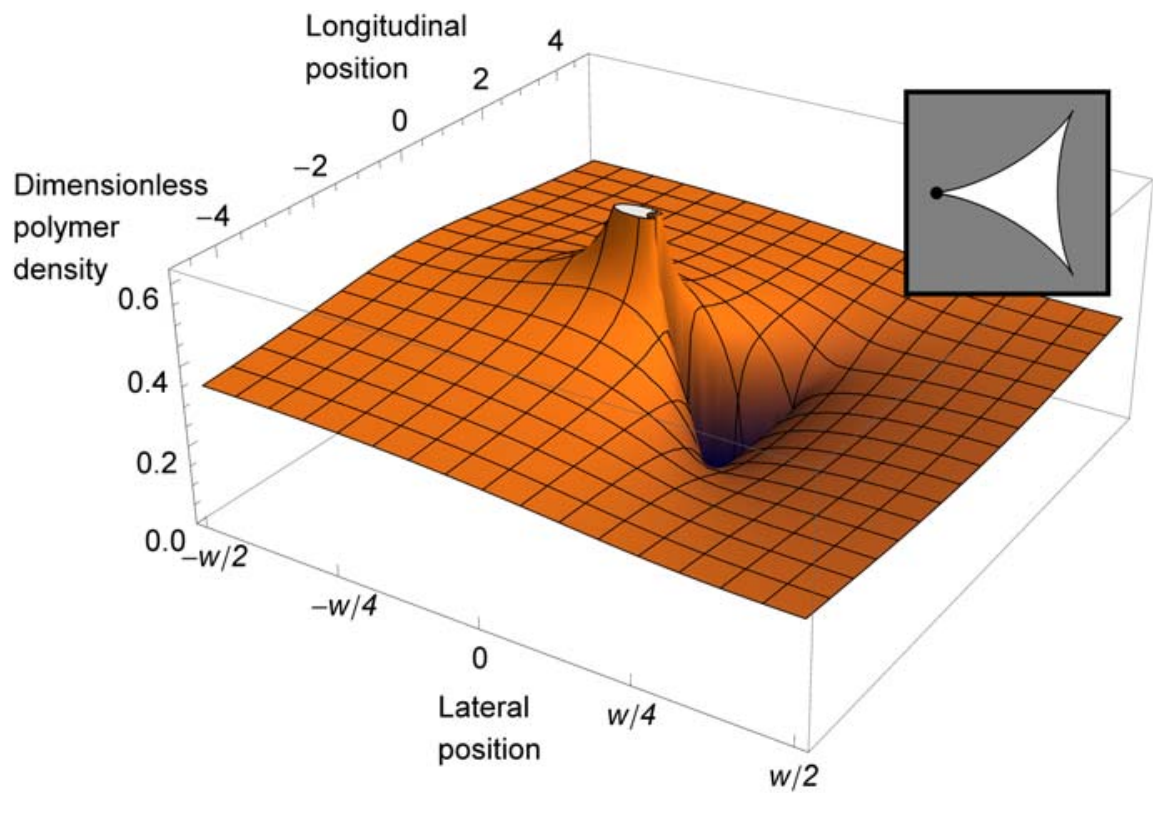}}
\caption{\label{phase}
Variation of the equilibrium directed-polymer density over the $(x,\tau)$ plane in the vicinity of a pin constraint.
The physical polymer density is the plotted quantity scaled by $N \pi/ \pwidth$.
Note that the gap (i.e., the region in which the present approximation scheme gives zero for the polymer density) is roughly triangular in shape, as shown by the unshaded area of the inset.  The divergence in the polymer density occurs only at precisely the location of the pin, depicted by a black dot in the inset. 
}
\label{fig:rhoxtplot}
\end{figure}

For a system of free fermions $\mathcal{E}(\rho)=(\hbar^2 \pi^2/6 m)\rho^2$, and the hydrodynamical equations may be expressed as the complex Hopf equation
\begin{subequations}
\begin{eqnarray}
\partial_t w - i w \partial_x w &=& 0,
\\
w(x,t) &\equiv& \frac{\hbar \, \pi}{m} \rho(x,t) + i v(x,t).
\end{eqnarray}
\end{subequations}
This equation can be solved numerically to describe the polymer density in the $x$-$\tau$ plane resulting from a pin, with the mapping between quantum and polymer parameters as in Eq.~(\ref{eq:mappings}).  
The previously obtained mean density profile serves as a boundary condition for the field $w$: $w(x,\tau_p)=(\hbar \, \pi/m) \, \mdensity(x)$ [$v(x,\tau_p) = 0$ due to symmetry considerations].
As the hydrodynamical equations are obtained by ignoring contributions associated with certain higher-gradient terms, they are expected to be quantitatively incorrect in the vicinity of any pins, because there the polymer density diverges (according to the mean-field type of approach that constitutes the bulk of the present paper), and it is higher-order terms that dominate and control such divergences. 
Indeed, for a broad class of systems, nonlinearities can lead to shock-wave behavior even for initially fairly smooth density profiles~\cite{abanov2}, and even more readily for the divergent behavior around a pin.
The polymer density resulting from a pin at $x=-0.06\pwidth$ with the partitioning $N_L=.74 N$ is shown in Fig.~\ref{fig:rhoxtplot}.
We note that the divergence in the polymer density at the pin immediately subsides
but the gap persists over a finite range in the longitudinal direction.
At large longitudinal distances, however, the polymer density profile returns to the uniform distribution of the pin-less system.

A qualitative analysis of large quantum fluctuations indicates that the probability $P(R)$ of a large fluctuation over a lengthscale $R$ has the form $P(R)\sim \exp(-\alpha R^2)$~\cite{abanov}.  This corresponds to our finding in the present paper that when the pin displacement is small (compared to the overall system size) and all polymers lie on one side of the pin the free-energy cost is quadratic in the displacement. However, when the finite temperature of such a quantum system is taken into account, the probability of a fluctuation becomes $P(R)\sim \exp(-\gamma R)$.  The analog of a finite temperature in the quantum system is a finite \emph{length} in the polymer system.  For a system in which the pin coordinate $\tau_p$ is located near enough to another system feature, such as another pin or an end of the system, the ground-state dominance approximation fails, and one instead finds that the free-energy cost of the pin would increase linearly with the displacement of the pin from its equilibrium position. Thus, a longitudinally short system, or one having longitudinally distributed pins, can display a super-Hookean response when a pin is displaced a small amount from its equilibrium position.

We remark that although the polymer system we have been exploring is formally equivalent to its quantum analog, the polymer system is more readily controllable. Large quantum fluctuations are, due to their rarity, difficult to observe. In contrast, the probability of occurrence
of the equivalent large thermal fluctuations of the polymer system can be measured indirectly, via the entropic force on a pin, which has the useful effect of forcing the system to assume what would otherwise be rare configurations.

\section{Concluding remarks}
\label{sec:conclusion}
Our central aim with this paper has been to develop a statistical-mechanical treatment of macroscopic systems of directed, two-dimensional, classical polymers in thermal equilibrium in the limit of infinitely strong repulsion between the polymers.  More specifically, our focus has been on the consequences of topological constraints---either point-like pins or spatially extended lines---that place limitations on the microscopic configurations accessible to the polymer systems.  These constraints have the effect of altering the equilibrium value of the spatial profile of the local polymer density and reducing the entropy (and, perforce, increase the free energy) of directed polymer systems and, when generically located, they give rise to forces that act on the constraints (and between them, if there are several of them).

The main technique that we employ to compute these effects is a mapping between two-dimensional systems of classical directed polymers in thermal equilibrium and the imaginary-time evolution of one-dimensional systems of quantal point-like particles in the ground state. (The particles can be taken to be either non-interacting identical fermions or identical bosons subject to infinitely repulsive short-ranged interactions.)\thinspace\  The determination of the alteration of statistical weight resulting from the presence of constraints for the polymer system is then ascertained via consideration of a rather unusual quantal amplitude: the matrix element of the imaginary-time evolution operator between the many-particle ground state and an appropriately partitioned configuration of the particle system.  This technique readily yields~\cite{largefluc} results for situations in which the constraints are located collinearly on a line that runs perpendicular to the direction preferred by the polymers.  By supplementing this technique with a quantum hydrodynamic approach we are able to establish the form of the polymer density not only along the aforementioned line on which the constraints lie but also at points away from this line.

Strikingly, in the limit of large polymer densities, in which fluctuations are small and a type of mean-field theory is accurate, we find that the effect of a point-like pin is to cause a divergent pile-up of the polymer density on the high-density side of the pin and a zero-density region (or gap) on the low-density side of the pin, the latter giving way to a continuously rising density beyond the gap.  
In addition, via the quantum hydrodynamic approach we find that the gap opened by the pin has a nonzero extent in the preferred direction, only gradually narrowing and closing.
  (Fluctuation corrections to this mean-field theory are expected to replace the zero-density region by a region of very small polymer-density.)\thinspace\
We also find for pins that are only mildly displaced from a zero-force (i.e., equilibrium) position the force acting on them is sub-Hookean, growing less than linearly with the displacement via a factor logarithmic in the displacement, and the gap created by them similarly grows sub-linearly with the displacement.  By contrast, for multiple pins that are separated from one another along the direction preferred by the polymers we find free energy costs that are super-Hookean, growing more than quadratically with the displacement.  One can regard these nonlinear responses as resulting from the effectively long-ranged interactions between polymer segments that emerge via short-ranged interactions between long polymer strands in regions that reach far from the segments in question.

In a forthcoming companion paper~\cite{ref:DZRcompanion} we shall discuss developments arising from the present work that encompass, inter alia, the impact of external potentials of various forms, the consequences of inter-polymer interactions that are finitely (as opposed to infinitely) strong or extended in range, the behavior of distinct polymer species within host polymer systems, and and the application of powerful techniques from quantum many-body physics, including bosonization and the Bethe Ansatz.

\begin{acknowledgments}
We thank Jennifer Curtis, Thierry Giamarchi, Sarang Gopalakrishnan, Michael Pustilnik and Andrew Zangwill for valuable discussions.
One of us (PMG) thanks for its hospitality the Aspen Center for Physics, where part of the work reported here was carried out.
This work was supported by grants NSF DMR 09-06780 (PMG), NSF PHY-1068511 (ST), DOE DEFG02-07ER46453 (DZR), the Alfred P. Sloan Foundation (ST),  and an NDSEG Fellowship  (DZR).
\end{acknowledgments}

\appendix

\section{$N$ fermion ground-state wave function}

Consider a system of $N$ noninteracting fermions moving freely one dimension on the line segment $0<x<\pi$ and subject to homogeneous (i.e., homogeneous Dirichlet) boundary conditions at $x=0$ and $x=\pi$.  The normalized wave functions associated with the single-particle energy eigenstates of such a system are given by
\begin{eqnarray}
\phi_j(x) = \sqrt{2/\pi} \sin (j x), \ \ j = 1, 2, \ldots.
\end{eqnarray}
In terms of these, the normalized $N$-particle ground-state wave function of the $N$-fermion system may be expressed in terms of a Slater determinant, as follows:
\begin{eqnarray}
\psi(x_1,\ldots,x_N) = \frac{1}{\sqrt{N!}}\det_{N\times N}\big[\phi_j(x_k)\big],
\end{eqnarray}
where $[\phi_j(x_k)]$ is the $N \times N$ matrix having $(j k)$ element $\phi_j(x_k)$.
Omitting $N$ factors of $\sqrt{2/\pi}$, the matrix $[ \phi_j(x_k) ]$ takes the form
\begin{eqnarray}
\left( \begin{array}{cccc}
\sin   x_1 & \sin   x_2  & \cdots & \sin   x_N \\
\sin 2 x_1 & \sin 2 x_2 & \cdots & \sin 2 x_N \\
\vdots & \vdots & \ddots & \vdots \\
\sin N x_1 & \sin N x_2 & \cdots & \sin N x_N
\end{array} \right).
\end{eqnarray}
To simplify the evaluation of the determinant of this matrix, we add linear combinations of the rows of the matrix to other rows, a procedure that leaves its value unchanged. Specifically, we make the replacements
\begin{eqnarray}
\phi_j(x_k) \rightarrow \phi_j(x_k) + \phi_{j-2}(x_k) - 2 \cos x_N \ \phi_{j-1}(x_k),
\nonumber
\end{eqnarray}
where we make the definition $\phi_j(x_k) = 0$ for $j<1$.  The resulting matrix is then given by
\begin{eqnarray}
\left( \begin{array}{c|c}
\sin   x_1 \ \cdots \ \sin x_{N-1} & \sin x_N
\\
\hline
\begin{matrix} 2(\cos x_k  - \cos x_N)
\\ \times \sin \left((j-1) x_k\right)
\end{matrix}
&
\begin{matrix} 0 \\
\vdots \\ 0
\end{matrix}
\end{array} \right).
\end{eqnarray}
Note, specifically, the vanishing of the elements in the lower $N-1$ rows in the $N^{\rm th}$
column. In evaluating the determinant of this matrix, we may extract $N-1$ factors of the form
$2(\cos x_k  - \cos x_N)$
from the rows in the lower left block of the matrix.
Thus, we obtain the recursive result, linking Slater determinants for $N$-particle and
$(N-1)$-particle systems:
\begin{eqnarray}
&&\det_{N \times N}
\big[\phi_j(x_k)\big]=
(-1)^{N-1} \sin x_N
\nonumber\\
&&
\quad\times
\prod_{k=1}^{N-1}
\left(2\cos x_k - 2\cos x_N\right)
\times
\!\!\!\!\!
 \det_{(N-1)\atop{\quad\times(N-1)}}
\!\!\!\!\!
\big[\phi_j(x_k)\big].
\end{eqnarray}
Next, we apply this relation recursively, and this arrive at the following, desired form for the $N$-particle ground-state wave function:
\begin{eqnarray}
&&\psi(x_1, x_2,\ldots,x_N)=
\frac{2^{N^2/2}}{\pi^{N/2}\sqrt{N!}}
\left(\prod\nolimits_{j = 1}^{N} \sin x_j\right)
\nonumber \\
&&\qquad\qquad\qquad\times
\prod\limits_{1 \le j < k \le N}
\big(\cos x_j - \cos x_k\big).
\end{eqnarray}
Note that for a system of hard-core bosons the corresponding ground-state wave function is simply the absolute value of this fermionic wave function.

\section{Elliptic integral representations of the pin constraint and free energy} \label{sec:ellipticals}

Given the form of the polymer density found in Sec.~\ref{sec:densityprofile} the pin constraint Eq.~(\ref{eq:pincon}) may be constructed in terms of the elliptic integral of the third kind as

\begin{subequations}
\begin{eqnarray} \label{eq:sgspnrelationship}
&&\frac{N_R}{N}=\frac{2(s_g-s_p)}{\pi \sqrt{(1-s_p)(1+s_g)}}  \\
&& \qquad \qquad \times \, \Pi \left( \frac{1+s_p}{1+s_g},\frac{(1+s_p)(1-s_g)}{(1-s_p)(1+s_g)} \right),  \nonumber \\
&& \Pi(n,m) \equiv \int_{0}^{1}\frac{d s}{1-n s^2} \frac{1}{\sqrt{(1-m s^2)(1-s^2)}}.
\end{eqnarray}
\end{subequations}

\noindent We may describe how $s_g$ varies with $s_p$ for a given partitioning by requiring that the partitioning given by $N_R$ is constant with respect to variations of $s_p$, so that

\begin{eqnarray}
\frac{d N_R}{d s_p} = \frac{\partial N_R}{\partial s_p} +\frac{\partial N_R}{\partial s_g}\frac{\partial s_g}{\partial s_p}=0.
\end{eqnarray}

\noindent By using Eq.~(\ref{eq:sgspnrelationship}), we thus find the result

\begin{eqnarray} \label{eq:spsgdifeq}
\frac{\partial s_g}{\partial s_p} = \frac{1+s_g}{1+s_p}\left(1-\frac{E\left(\frac{(1+s_p)(1-s_g)}{(1-s_p)(1+s_g)}\right)
}{K\left(\frac{(1+s_p)(1-s_g)}{(1-s_p)(1+s_g)}\right)}\right),
\end{eqnarray}

\noindent where $K$ and $E$ are elliptic integrals of the first and second kind, respectively:

\begin{subequations}
\begin{eqnarray}
K(k) &\equiv& \int_{0}^{1}  \frac{d s}{\sqrt{(1-k^2 s^2)(1-s^2)}},  \\
E(k) &\equiv& \int_{0}^{1} d s \ \frac{\sqrt{1-k^2 s^2}}{\sqrt{1-s^2}}.
\end{eqnarray}
\end{subequations}

For a pin infinitesimally displaced from its equilibrium position, so that $s_p-s_g \approx \gap \cos x_p$ 
 with $\gap \ll 1$, the right-hand side version of the pin constraint~(\ref{eq:pincon}), may similarly be asymptotically expanded in $\gap$ to obtain the condition

\begin{eqnarray}
\pi \frac{N_R}{N} \approx \int_{s_g}^{1} \frac{d s}{\sqrt{1-s^2}} \left[ 1 + \frac{\gap \cos x_p}{2}\frac{1}{s-s_p}\right]
\nonumber \\
\approx  \frac{\pi}{2} - x_g + \frac{\gap}{2} \ln \left[ \frac{2 \cos^2 x_p}{|\gap|}\right].
\end{eqnarray}

\noindent This equation yields the asymptotic relation between gap size and displacement given in Eq.~(\ref{eq:smallgap}).

\begin{figure}[hh]
\centerline{
\includegraphics[width=.48\textwidth]{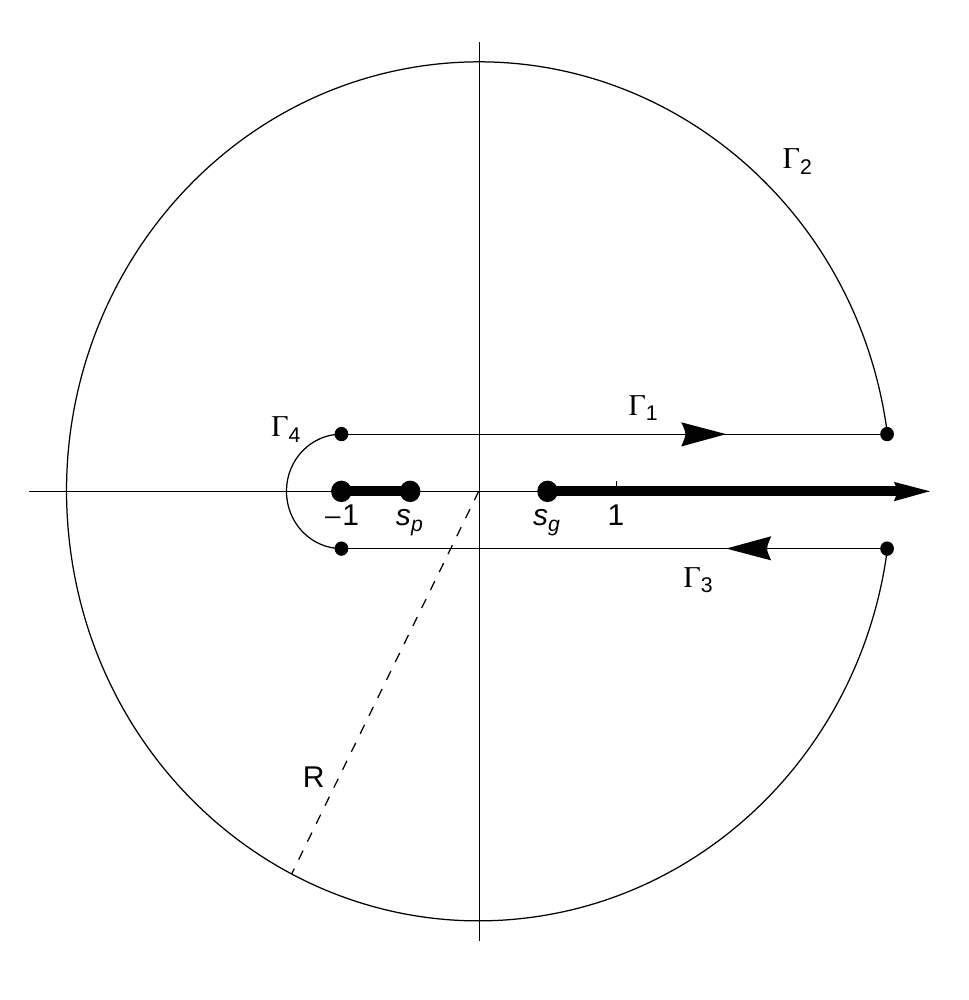}}
\caption{
The contour $\Gamma$ used in Eq.~(\ref{eq:contourequation}). Branch cuts run from $s=-1$ to $s=\min(s_p,s_g)$ and from 
 $s=\max(s_p,s_g)$ to $s=+\infty$. The arc $\Gamma_4$ has a radius $\epsilon$ which will be taken to zero, and the arc $\Gamma_2$ has a radius $R$ which will be taken to infinity.
}
\label{fig:contour}
\end{figure}

We turn our attention now to the free energy cost of the pin, Eq.~(\ref{eq:FEincrease}), which can be written in terms of the scaled effective potentials $\phi_L$ and $\phi_R$ experienced by polymers on the left and right side of the pin, respectively, as 

\begin{subequations}
\begin{eqnarray} \label{eq:nrphir}
\frac{\Delta \mathcal{F}}{N^2 T} &=& \frac{N_L}{N} \phi_L + \frac{N_R}{N} \phi_R, \\
\phi_L & \equiv & \int_{-1}^{1} d s \, \ln\left[2(1+s)\right] Q(s),  \\
\phi_R & \equiv & \int_{-1}^{1} d s \, \ln\left[2(1-s)\right] Q(s), 
\end{eqnarray}
\end{subequations}

\noindent with the transformed coordinate $s$ and polymer density $Q(s)$ defined in the main text; see Eq.~(\ref{eq:newvar}).
As we shall see, it is useful to represent $\phi_R$ as an elliptic integral, and to do this it is necessary to obtain a form without the logarithm.
To this end, we extend $s$ into the complex plane, and make use of the residue theory result 

\begin{eqnarray} \label{eq:contourequation}
\oint_\Gamma d s \ln (1-s) Q(s) = 0,
\end{eqnarray}

\noindent where the keyhole contour $\Gamma$ comprises the segments $\Gamma_1, \ldots, \Gamma_4$ as shown in Fig.~\ref{fig:contour}.
Now, the integrand has the form

\begin{widetext}

\begin{subequations}
\begin{eqnarray}
\ln(1-s) \, Q(s)&=&
\frac{1}{\pi} \Big(\ln |1 - s| + i \arg(1-s)\Big)\sqrt{\left| \frac{s-s_g}{(1-s^2)(s-s_p)}\right|} \, e^{i \phi(s)},  \\
\phi(s) &\equiv& \big (\arg(s-s_g)-\arg(1-s) 
-\arg(1+s)-\arg(s-s_p) \big )/2, 
\end{eqnarray}
\end{subequations}

\noindent which is analytic in the complex plane except on the branch cuts. Note that although the physical polymer density is zero in the gap, we are allowing $Q(s)$ to be nonzero but imaginary in the gap, and complex in the complex $s$ plane. We choose the branch cuts to run from $s=-1$ to $s=\min(s_p,s_g)$ and from 
 $s=\max(s_p,s_g)$ to $s=+\infty$, as shown in Fig.~\ref{fig:contour}.

Whereas the integral along the inner contour $\Gamma_4$ vanishes as its radius $\epsilon$ goes to zero, the integral along the outer contour $\Gamma_2$ (i.e., along $|s|=R$), does not vanish. In the limit $R \gg 1$ this contour integral becomes

\begin{eqnarray}
\int_{\Gamma_2} d s \ln(1-s) \, Q(s) \approx 
\int_{0}^{2 \pi} \left(i R e^{i \theta} d \theta \right)
\Big[\ln R + i \left(\theta-\pi\right)\Big]\frac{i}{\pi R e^{i \theta}}
=-2 \ln R.
\end{eqnarray}

\noindent Thus, the contour integral of Eq.~(\ref{eq:contourequation}) yields

\begin{eqnarray} \label{eq:contourequation}
2 \ln R \approx \int_{-1}^{1} d s \ln|1-s| \, Q(s+i \epsilon)
+\int_{1}^{R} d s \Big[\ln|1-s|-\pi i\Big] \, Q(s+i \epsilon) \nonumber \\
+\int_{R}^{1} d s \Big[\ln|1-s|+\pi i\Big] \, Q(s-i \epsilon)
+\int_{1}^{-1} d s \ln|1-s| \, Q(s-i \epsilon).
\end{eqnarray}

\end{widetext}

\noindent Elementary cancellations among parts of these integrals occur, so that upon taking the limit $R\rightarrow \infty$, Eq.~(\ref{eq:contourequation} (which is exact in this limit) yields

\begin{subequations}
\begin{eqnarray}
\phi_R &=& \int_{1}^{\infty} d s \left[\frac{1}{s} - \sqrt{ \frac{s-s_g}{(s^2-1)(s-s_p)}}\right],  \\
\phi_L &=& \int_{1}^{\infty} d s \left[\frac{1}{s} - \sqrt{ \frac{s+s_g}{(s^2-1)(s+s_p)}}\right].
\end{eqnarray}
\end{subequations}

\noindent Using these representations of $\phi_L$ and $\phi_R$, the free energy may be expressed in terms of elliptic integrals, although the complete expression is rather complicated. 
Strikingly, however, a simple result follows for the force on the pin, $-d \mathcal{F}/d x_p$. In particular, by differentiating Eq.~(\ref{eq:nrphir}) we find

\begin{eqnarray}
\frac{1}{N^2 T} \frac{d \mathcal{F}}{d s_p}= \frac{N_L}{N} \frac{d \phi_L}{d s_p} + \frac{N_R}{N} \frac{d \phi_R}{d s_p},
\end{eqnarray}

where 

\begin{eqnarray}
\frac{d \phi_{L/R}}{d s_p} = \frac{\partial \phi_{L/R}}{\partial s_p} + \frac{\partial s_g}{\partial s_p}
\frac{\partial \phi_{L/R}}{\partial s_g}.
\end{eqnarray}

\noindent Thus, by consideration of the elliptic integral representation, it can be shown that

\begin{eqnarray}
\frac{1}{N^2 T}\frac{d \mathcal{F}}{d s_p} =\frac{s_p-s_g}{1-s_p^2},
\end{eqnarray}

\noindent and from this result follows the force on the pin given in Eq.~(\ref{eq:fedifeq}).


\begin{thebibliography}{99}
\bibitem{knotting} 
Real polymer systems are of course three dimensional, with polymers able to crossing either over or under one another by exploiting the third dimension.  Thus, each particular configuration of a quasi-two-dimensional directed-polymer system corresponds to an element of the braid group.  If polymer endpoints are not free to fluctuate then the thermal ensemble would be restricted to a particular sector of the braid group of fixed topology.  As we shall discuss in a companion paper, this effect of topology conservation can give rise to an effective contact repulsion between monomers and to glassiness.  In the present paper we avoid these issues by focusing on strictly non-crossing polymers, which are automatically restricted to the trivial braiding.
\bibitem{gibbsdenominator}
If the polymers are indistinguishable, both by their composition and by their boundary conditions, a Gibbs denominator of $N!$ must be included to ensure that the partition function be extensive. Such a change would have no effect on the observables considered here.
\bibitem{fermistates}
If we choose periodic boundary conditions $P^i(\{X_n\})$=$P^f(\{X_n\})$ we may then associate this shared probability with $|\Psi(\{X_n\})|^2$. This would still not permit the general use of fermionic boundary conditions, which would exclude from our ensemble permitted polymer crossings.
\bibitem{largefluc}
From this perspective, we see that directed polymer systems provide access to phenomena that, in the setting of QMB physics, amount to large, and therefore rare, fluctuations.
\end{thebibliography}
\end{document}